\documentclass{article}

\usepackage[english]{babel}
\usepackage[ansinew]{inputenc}

\usepackage{amsfonts}
\usepackage{amssymb}
\usepackage{subfigure}

\usepackage{psfrag,graphicx}
\textheight \paperheight
\textwidth \paperwidth
\topskip \baselineskip
\newcommand{\app}[3]{Acta Phys. Pol. \textbf{#1} ({#2}) {#3}}
\newcommand{\nucl}[3]{Nucl. Phys. \textbf{#1} ({#2}) {#3}}

\newcommand{\prd}[3]{Phys. Rev. \textbf{D{#1}} ({#2}) {#3}}
\newcommand{\prev}[3]{Phys. Rev. \textbf{{#1}} ({#2}) {#3}}
\newcommand{\plb}[3]{Phys. Lett. \textbf{B{#1}} ({#2}) {#3}}
\newcommand{\prep}[3]{Phys. Rep. \textbf{{#1}} ({#2}) {#3}}

\begin{document}


\title{Gribov copies, Lattice QCD and the gluon propagator}

\author{P. J. Silva\thanks{email: psilva@teor.fis.uc.pt}, 
        O. Oliveira\thanks{email: orlando@teor.fis.uc.pt} \\
        Centro de Física Computacional \\ 
        Departamento de Física \\  
        Universidade de Coimbra \\
        3004 - 516 Coimbra, Portugal}

\maketitle

\begin{abstract}
We address the problem of Gribov copies in lattice QCD. The gluon propagator
is computed, in the Landau gauge, using 302 ($\beta = 5.8$)
$12^4$ configurations gauge fixed to different copies. The results of the
simulation shows that: i) the effect of Gribov copies is small (less than
10\%); ii) Gribov copies change essentially the lowest momenta components 
($q < 2.6$ GeV); iii) within the statistical accuracy of our simulation, the
effect of Gribov copies is resolved if statistical errors are multiplied by a
factor of two or three. Moreover, when modelling the gluon propagator, 
different sets of Gribov copies produce different sets of parameters not, 
necessarily, compatible within one standard deviation. Finally, our  data 
supports a gluon propagator which, for large momenta, behaves like a
massive gluon propagator with a mass of $1.1$ GeV.
\end{abstract}

\textit{keywords}: 
Lattice QCD, Landau Gauge, Gauge Fixing, Gribov Copies, Gluon Propagator,
Gluon Mass

\textit{PACS} 12.38.G, 11.15.H

\section{Introduction and Motivation}

Quantum Chromodynamics (QCD) is the theory that describes the interaction 
between quarks and gluons. The definition of the QCD generating functional
\textit{à la} Faddeev-Popov \cite{feyman,dewitt,faddeev_popov} requires a 
choice of a gauge condition, uniquely satisfied in each gauge orbit, i.e. on 
each set of fields related by a gauge transformation. 
For the Landau, the Coulomb gauge and for small field amplitudes, the gauge 
condition is uniquely satisfied in each gauge orbit. However, if large field 
amplitudes are involved, the gauge fixing condition has multiple solutions in 
each gauge orbit \cite{gribov,sciuto}, the Gribov copies. 

Gribov copies appear when large field amplitudes are involved and rise 
the question of how to define the generating functional for the 
nonperturbative regime of Quantum Chromodynamics. Moreover, in \cite{singer} it
was proved that it is not possible to find a local continuous and
unambigous gauge fixing condition for any SU(N) gauge theory defined on the
manifold $S_4$. A similar result for the four-torus was obtained
in \cite{killingback}.

For the continuum formulation of QCD, Zwanziger argued in \cite{Zwanziger03}
that the Landau gauge Faddeev-Popov formula
\begin{equation}
\delta ( \partial A) ~ \det [ - \partial \cdot D(A) ] ~ 
\exp [ - S_{YM} (A) ] ~ ,
\end{equation}
restricted to the region where the Faddeev-Popov operator is positive definite
$- \partial \cdot D(A) > 0$, the Gribov region $\Omega$, provides an exact 
non-perturbative quantisation for QCD. This result helps to eliminate some 
theoretical questions about the investigations of QCD using Dyson-Schwinger 
equations (DSE). Nevertheless, in what concerns the non-perturbative regime of
QCD, being unable to solve exactly the DSE, the results of such studies should
be compared to lattice results. In this way one can test the validity of the
approximations and ansatz used to solve the DSE and, simultaneously, 
the lattice algorithms.

The formulation of gauge theories on the lattice does not require gauge 
fixing. As long as one is interested only on gauge invariant 
operators, the lattice calculation is not plagued with the problem of Gribov
copies. However, the investigation of the Green's functions of the 
fundamental fields, such as the gluon, ghost and fermion propagators, implies
the choice of a gauge. On the lattice, typically, a simulation begins by 
generating a number of thermalized gauge configurations. In order to compute,
for example, the propagators, each configuration is then rotated to satisfy a 
given gauge fixing condition. Finally, the propagator is computed using these 
rotated configurations. For the Landau gauge, gauge fixing is implemented by 
computing a maximum of a given function defined on the gauge orbits. Now, the 
problem of the Gribov copies is due to the several maxima of the 
maximizing function. The first observations and studies of lattice Gribov 
copies were done long ago 
\cite{nakamura,forcrand91,marinari,Pa92,MaRo93,Pa94}. However, how the choice
of Gribov copies change the correlation functions is not yet clear.

On the continuum formulation, Gribov \cite{gribov} studied SU(2) gauge theory.
His purposal to solve the problem of the different copies was to restrict the 
functional integration space to the so called Gribov region $\Omega$. The 
gluon propagator computed by functional integrating the gluon fields over
$\Omega$ does not show the usual perturbative $1/q^2$ behaviour but, instead,
$q^2/(q^4 + M^4)$, with $M$ being a mass scale which measures the volume of 
$\Omega$. Note that the two propagators agree for the high energy regime. 

On the lattice, there was a number of studies about Gribov copies and
different observables in various gauges. In this paper we will be mainly
concerned about the gluon propagator computed in the Landau gauge.
For a general discussion about lattice Gribov copies see, for example, 
\cite{giusti} and references therein. For the SU(2) group, the gauge and ghost
propagators versus Gribov copies were studied in \cite{cucchieri97,FuNa03}. 
The authors claim that the gluon propagator is not sensible to Gribov copies
in the weak coupling regime\footnote{Note that, in the strong coupling regime,
Cucchieri is able to see differences on the propagator due to Gribov copies.}. 
For the ghost propagator, the simulations performed by the first author shows
that, close to the continuum, the propagator is again not sensible to Gribov
copies. In the second study, it is claimed a reduction of 6\% for the central
value of the smaller momenta ghost propagator and a reduction of 4\% on the
Kugo-Ojima parameter\footnote{See, also, \cite{Na03}.}. The SU(2) simulations 
suggest that the influence of Gribov copies is at the level of the simulation 
statistical error. For SU(3) there is no systematic study but it is believed 
that the Gribov noise is contained within the statistical error of the Monte 
Carlo.

In this paper we study the pure gauge lattice QCD gluon propagator in the
Landau gauge and try to understand the role of the Gribov copies. 
We compute the gluon propagator for
302 configurations, with $\beta = 5.8$, for a lattice size of $12^4$. 
Our results show that, although being a small effect (less than 6-10\%), 
the Gribov copies change the lowest momenta components of the gluon 
propagator. This effect 
is illustrated fitting the gluon propagator and comparing the results for sets
of configurations built from different copies. Gribov copies influence can go
from a doubling of the statistical error, to the extreme case of changing the
functional form of the propagator.

The paper is organised as follows. Section II sets the field definitions
and notation used in this work. In section III, the Landau gauge is discussed,
both on the continuum and in lattice QCD. Moreover, the algorithm used here
is sketched. In section IV, the results for the role of Gribov copies in
the gluon propagator are reported. Finally, in section V our results are
discussed.

\section{Field Definitions and Notation}

In the lattice formulation of QCD, the gluon fields $A^a_\mu$ are replaced by 
the links
\begin{equation}
  U_\mu (x) ~ = ~ e^{ i a g_0 A_\mu (x + a \hat{e}_\mu / 2) }
  ~ + ~ \mathcal{O}( a^3 ) 
 ~ ~ ~ \in ~ SU(3)  \, ,
\end{equation}
where $\hat{e}_\mu$ are unit vectors along $\mu$ direction. QCD is a gauge 
theory, therefore the fields related by gauge transformations
\begin{equation}
  U_\mu (x) ~ \longrightarrow ~ g(x) ~ U_\mu (x) ~ 
 g^\dagger (x + a \hat{e}_\mu) \, ,
 \hspace{1cm} g \in SU(3) \, ,
\end{equation}
are physically equivalent. The set of links related by gauge transformations
to $U_\mu (x)$ is the orbit of $U_\mu (x)$. 

The gluon field associated to a gauge configuration is given by
\begin{equation}
A_\mu (x + a \hat{e}_\mu / 2) ~ = ~
   \frac{1}{2 i g_0} \Big[ U_\mu (x)  -  U^\dagger_\mu (x) \Big] ~ - ~
   \frac{1}{6 i g_0} \mbox{Tr}
         \Big[ U_\mu (x)  -  U^\dagger_\mu (x) \Big] 
\end{equation}
up to corrections of order $a^2$.

On the lattice, due to the periodic boundary conditions, the discrete momenta
available are
\begin{equation}
  \hat{q}_\mu ~ = ~ \frac{2 \pi n_\mu}{a L_\mu} \, ,
  \hspace{0.5cm}  n_\mu ~ = ~ 0, \, 1, \, \dots \, L_\mu -1 \, ,
\end{equation}
where $L_\mu$ is the lattice length over direction $\mu$. The momentum
space link is
\begin{equation}
   U_\mu ( \hat{q} ) ~ = ~ \sum \limits_x ~ e^{-i \hat{q} x} ~
         U_\mu (x)
\end{equation}
and the momentum space gluon field
\begin{eqnarray}
   A_\mu ( \hat{q} ) & = & \sum \limits_x ~ 
         e^{-i \hat{q} \left(x + a \hat{e}_\mu / 2\right)} ~
         A_\mu (x + a \hat{e}_\mu / 2 ) \nonumber \\
         & = & \frac{e^{-i \hat{q}_\mu a/2}}{2 i g_0}
               \Bigg\{ ~
                   \Big[ U_\mu ( \hat{q} )  -  U^\dagger_\mu (- \hat{q}) \Big]
                   \, - \,
                   \frac{1}{3} \mbox{Tr}
                       \Big[ U_\mu ( \hat{q} )  -  U^\dagger_\mu (- \hat{q} )
 \Big] ~
               \Bigg\} ~ .
\end{eqnarray}

The gluon propagator is the gluon two point correlation function. The 
dimensionless lattice two point function is
\begin{equation}
 \langle A^a_\mu (\hat{q}) ~ A^b_\nu (\hat{q}')  \rangle ~  = ~
  D^{ab}_{\mu\nu} ( \hat{q} ) ~ V ~ \delta( \hat{q} + \hat{q}' ) ~ .  
\end{equation}
On the continuum, the momentum space propagator in the Landau gauge is
given by
\begin{equation}
D^{ab}_{\mu\nu} ( \hat{q} ) ~ = ~ \delta^{ab} ~
   \Big( \delta_{\mu\nu} ~ - ~ \frac{q_\mu q_\nu}{q^2} \Big) ~
   D( q^2 ) ~ . \label{propcont}
\end{equation}
Assuming that the deviations from the continuum are negligable, the lattice
scalar function $D( q^2 )$ can be computed directly from (\ref{propcont}) as 
follows
\begin{equation}
  D( q^2 ) ~ = ~ \frac{2}{(N^2_c-1)(N_d-1) V} \sum\limits_{\mu} ~ 
                 \langle ~ \mbox{Tr} \left[  A_\mu (\hat{q}) \, 
                                           A_\mu ( -\hat{q} ) \right] ~ \rangle
      \, ,
      \hspace{0.3cm} q \ne 0 \, ,
 \label{Dq2}
\end{equation}
and
\begin{equation}
  D( 0 ) ~ = ~     \frac{2}{(N^2_c-1) N_d V} \sum\limits_{\mu} ~ 
                 \langle ~ \mbox{Tr} \left[  A_\mu (\hat{q}) \, 
                                           A_\mu ( -\hat{q} ) \right] ~ \rangle  \, , \hspace{0.3cm}
       q = 0 \, ,
    \label{Dq20}
\end{equation}
where
\begin{equation}
  q_\mu \, = \, \frac{2}{a} ~ \sin \Big( \frac{\hat{q}_\mu a}{2} \Big) \, ,
\end{equation}
$N_c = 3$ is the dimension of the group, $N_d = 4$ the number of spacetime
dimensions and $V$ is the lattice volume.

\section{The Landau Gauge}

\subsection{The Continuum Landau Gauge}

On the continuum, the Landau gauge is defined by
\begin{equation}
  \partial_\mu A_\mu ~ = ~ 0 \, .\label{landau_cont}
\end{equation}
This condition defines the hyperplane of transverse configurations
\begin{equation}
\Gamma ~ \equiv ~ \{A: ~ \partial \cdot A \, = \, 0 \} ~ .
\end{equation}
It is well known \cite{gribov} that $\Gamma$ includes more than one
configuration from each gauge orbit. In order to try to solve the
problem of the nonperturbative gauge fixing, Gribov suggested the use of
additional conditions, namely the restriction of physical configurational 
space to the region
\begin{equation}
   \Omega  ~ \equiv  ~ \{ A:~ \partial\cdot A \, = \, 0,~ 
                              \textit{M}[A] \, \geq \, 0 \} ~ \subset ~ \Gamma
 \, ,
\end{equation}
where $\textit{M}[A] ~ \equiv ~ - \nabla \cdot D[A] $ is the Faddeev-Popov
operator. However, $\Omega$ is not free of Gribov copies and does not provide 
a proper definition of physical configurations.

A suitable definition of the physical configurational space is given by the
fundamental modular region $\Lambda  \subset  \Omega$, the set of the absolute 
minima of the functional
\begin{equation}
   F_{A}[g] ~ =  ~ \int d^{4}x ~ \sum_{\mu} ~ 
    \mbox{Tr}\left[A_{\mu}^{g}(x)A_{\mu}^{g}(x)\right] \, .
 \label{fcont}
\end{equation}
The fundamental modular region $\Lambda$ is a convex manifold \cite{semyonov}
and each gauge orbit intersects the interior of $\Lambda$ only once
\cite{GAnt91,baal92}, i.e. its interior consists of non-degenerate
absolute minima. On the boundary $\partial \Lambda$ there are degenerate 
absolute minima, i.e. different boundary points are Gribov copies of each 
other \cite{baal92,baal94,baal95}. The interior of $\Lambda$, the region of
absolute minima of (\ref{fcont}),  identifies a region free of Gribov 
copies.

\subsection{The Lattice Landau Gauge}

On the lattice, the situation is similar to the continuum theory
\cite{zwanziger92,zwanziger94,cucc9711024}. The interior of $\Lambda$ consists
of non-degenerate absolute minima of the lattice version of (\ref{fcont})
and Gribov copies can occur at the boundary $\partial \Lambda$. However,
for a finite lattice, the boundary $\partial \Lambda$, where degenerate
minima may occur, has zero measure and the presence of these minima can be
ignored \cite{zwanziger94}.

On the lattice, the Landau gauge is defined by maximising the functional
\begin{equation}
   F_{U}[g] ~ = ~ C_{F}\sum_{x,\mu} \, \mbox{Re} \, \{ \, \mbox{Tr} \,
       [ g(x)U_{\mu}(x)g^{\dagger }(x+\hat{\mu}) ] \, \}  \label{f}
\end{equation}
where 
\begin{equation}
   C_{F}  ~ = ~ \frac{1}{N_{d}N_{c}V}
\end{equation}
is a normalization constant.  Let $U_\mu$ be the configuration that maximises
$F_U[g]$ on a given gauge orbit. For configurations near $U_\mu$ on its gauge
orbit, we have
\begin{eqnarray}
 F_U [ 1 + i \omega (x)]  =  F_U [ 1 ]  +  
            \frac{C_F}{4} \sum_{x,\mu}  i \omega^a (x) 
            \mbox{Tr} \Big[ & &
                      \lambda^a \left( U_\mu (x) \, - \,
                                         U_\mu ( x - \hat{\mu}) \right)
                            ~ -  \nonumber \\
        & &
                            \lambda^a \left( U^\dagger_\mu (x) \, - \,
                                              U^\dagger_\mu ( x - \hat{\mu}) 
                                    \right)
\Big] \, ,
\end{eqnarray}
where $\lambda^a$ are the Gell-Mann matrices. By definition, $U_\mu$ is a 
stationary point of $F$, therefore 
\begin{eqnarray}
\frac{\partial F}{\partial \omega^a (x)} ~ = ~
\frac{i \, C_F}{4} \sum_{\mu} 
            \mbox{Tr} \Big[ & &
                      \lambda^a \left( U_\mu (x) \, - \,
                                         U_\mu ( x - \hat{\mu}) \right)
                            ~ -  \nonumber \\
        & &
                            \lambda^a \left( U^\dagger_\mu (x) \, - \,
                                              U^\dagger_\mu ( x - \hat{\mu}) 
                                    \right)
\Big] ~ = ~ 0 \, .
 \label{statf}
\end{eqnarray}
In terms of the gluon field, this condition reads
\begin{equation}
  \sum_{\mu} \mbox{Tr} \Big[ ~ \lambda^a 
                               \left( A_{\mu}(x + a \hat{\mu}/2) - 
                                      A_{\mu}(x - a \hat{\mu}/2)
                               \right) ~\Big] \, + \, 
   \mathcal{O} (a^2) ~ =  ~ 0 \, ,
\end{equation}
or
\begin{equation}
  \sum_{\mu} \partial_\mu A^a_{\mu}(x) ~ + ~
   \mathcal{O} (a) ~ =  ~ 0 \, ,
\end{equation}
i.e. (\ref{statf}) is the lattice equivalent of the continuum Landau gauge 
condition. The lattice Faddeev-Popov operator $M(U)$  is given by the second 
derivative of (\ref{f}).

Similarly to the continuum theory, on the lattice one defines the region of
stationary points  of (\ref{f})
\begin{equation}
   \Gamma ~ \equiv ~ \{U: ~ \partial \cdot A(U) \, = \, 0 \} \, ,
\end{equation}
the Gribov's region $\Omega$ of the maxima of (\ref{f}),
\begin{equation}
   \Omega ~ \equiv ~ \{U: ~ \partial \cdot A(U)=0 ~ \mbox{and} ~ M(U) \geq 0 \}
\end{equation}
and the fundamental modular region $\Lambda$ defined as the set of the 
absolute maxima of (\ref{f}). 

A proper definition of the lattice Landau gauge chooses from each gauge
orbit, the configuration belonging to the interior of $\Lambda$.

\subsection{Gauge Fixing Algorithm}

On the lattice, gauge fixing is implemented by maximizing $F_U [g]$. In this 
work, the gauge fixing algorithm used is a Fourier accelerated steepest 
descent method (SD) as defined in \cite{davies}. In each iteration, the
algorithm chooses
\begin{equation}
   g(x) \, = \, \exp \Bigg[ \hat{F}^{-1} \,
                               \frac{\alpha}{2} \,
                               \frac{ p_{max}^{2} a^{2}}{ p^{2} a^{2} } \,
                               \hat{F} \, 
           \left( \sum_\nu  \Delta_{- \nu} \left[ U_\nu ( x ) \, - \,
                                          U^\dagger_\nu ( x ) \right]
                            \, - \, \mbox{trace} \right) \Bigg]
\label{sd}
\end{equation}
where
\begin{equation}
\Delta_{-\nu} \left( U_\mu (x) \right) ~ =  ~ 
       U_\mu ( x - a \hat{e}_\nu) \, - \, U_\mu ( x ) \, ,
\label{delta}
\end{equation}
$p^{2}$ are the eigenvalues of $(-\partial^{2})$, $a$ is the lattice spacing
and $\hat{F}$ represents a fast Fourier transform (FFT). For the parameter
$\alpha$ we use the value 0.08 \cite{davies}. For numerical purposes, it is
enough to expand to first order the exponential in (\ref{sd}), followed by a 
reunitarization of $g(x)$.

On the gauge fixing process, the quality of the gauge fixing is measured
by
\begin{equation}
  \theta ~ = ~ \frac{1}{VN_{c}} ~ \sum_{x} \mbox{Tr}
      [\Delta(x)\Delta^{\dag}(x)] \label{theta}
\end{equation}
where 
\begin{equation}
 \Delta(x) ~ = ~ \sum_{\nu} \left[ U_\nu ( x - a \hat{e}_\nu) \, - \,
                                   U^\dagger_\nu (x) \, - \, \mbox{h.c.}
                                   \, - \, \mbox{trace} \right]
\end{equation}
is the lattice version of $\partial_\mu A_\mu \, = \, 0$.

\section{The Gluon Propagator}

In this work only pure gauge quenched configurations are considered. The
Wilson action configurations were generated with version 6 of MILC code 
\cite{milc}. 

The function $F_U$ has many maxima - see, for example, \cite{nosso}. 
In each gauge orbit, the different maxima are different configurations and, 
therefore, the gluon propagator changes according to the chosen set of maxima.
In order to study such dependence, 302 gauge configurations were generated for
a $12^4$ lattice and for $\beta = 5.8$, using a combined update of 
4 over-relaxed and 5 quasi-heat bath Cabbibo-Mariani updates, with a 
separation between configurations of 3000 combined updates. 
To each gauge configuration, 500 SD gauge fixings, starting 
from different randomly chosen points, were performed requiring
\begin{equation}
 \theta ~ = ~ \frac{1}{V N_c} \sum\limits_{x} 
              \mbox{Tr} \left[ \Delta(x) \Delta^\dagger (x) \right] ~ = ~
       \frac{1}{V N_c} \sum\limits_{x} \left| \partial \cdot A \right|^2 
       ~ < ~ 10^{-15}  ~ .
\end{equation}
From these 500 SD, on each gauge orbit, we keep the gauge configurations 
associated to the largest maximum of $F_U$ (named MAX in the following), the 
smallest maximum of $F_U$ (named MIN) and three random values of $F_U$
(RND1, RND2, RND3), generated starting the gauge fixing process by choosing
always the same random $g(x)$ matrices. A further gauge fixing (named ID), 
starting the gauge fixing process by setting all $g(x) ~ = ~ 1$, was performed
to all gauge configurations. Another gauge fixing (named RND), starting the
gauge fixing process by choosing always the same random $g(x)$ matrices, was
performed to all configurations.

\begin{center}
\begin{figure}[t]
  \subfigure[Scalar Function as function of $\hat{q}$.]{\label{Fig_Dq2_qhat}
  \begin{minipage}[b]{\textwidth}
    \psfrag{EIXO X}{$\hat{q}$}
    \psfrag{EIXO Y}{$D(q^2)$}
    \psfrag{TITULO}{}
    \centering
    \includegraphics[origin=c,scale=0.45]{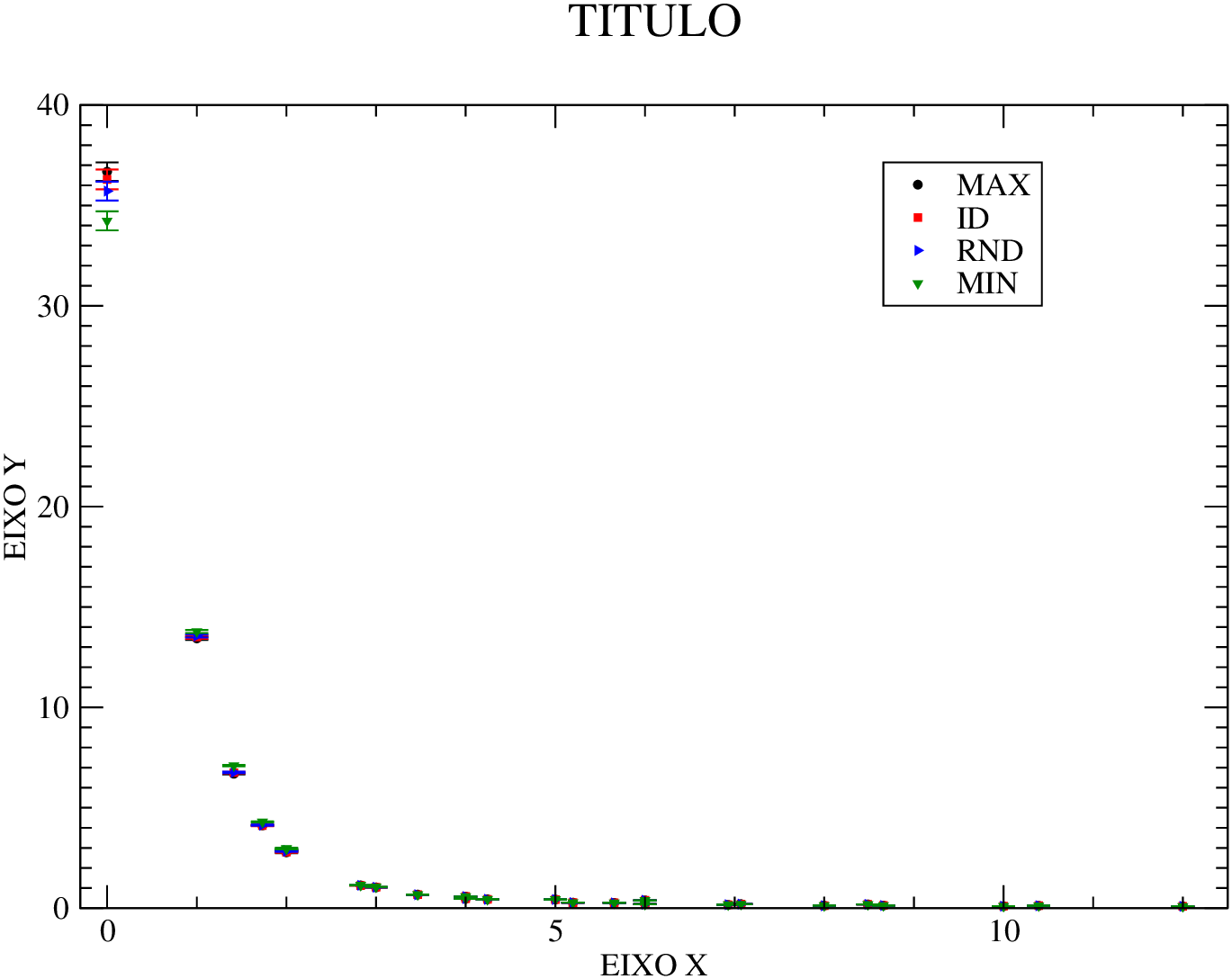}
  \end{minipage} } \\

  \subfigure[Scalar Function as function of $q$.]{ \label{Fig_Dq2_q}
  \begin{minipage}[b]{\textwidth}
    \psfrag{EIXO X}{$q$}
    \psfrag{EIXO Y}{$D(q^2)$}
    \psfrag{TITULO}{}
    \centering
    \includegraphics[origin=c,scale=0.45]{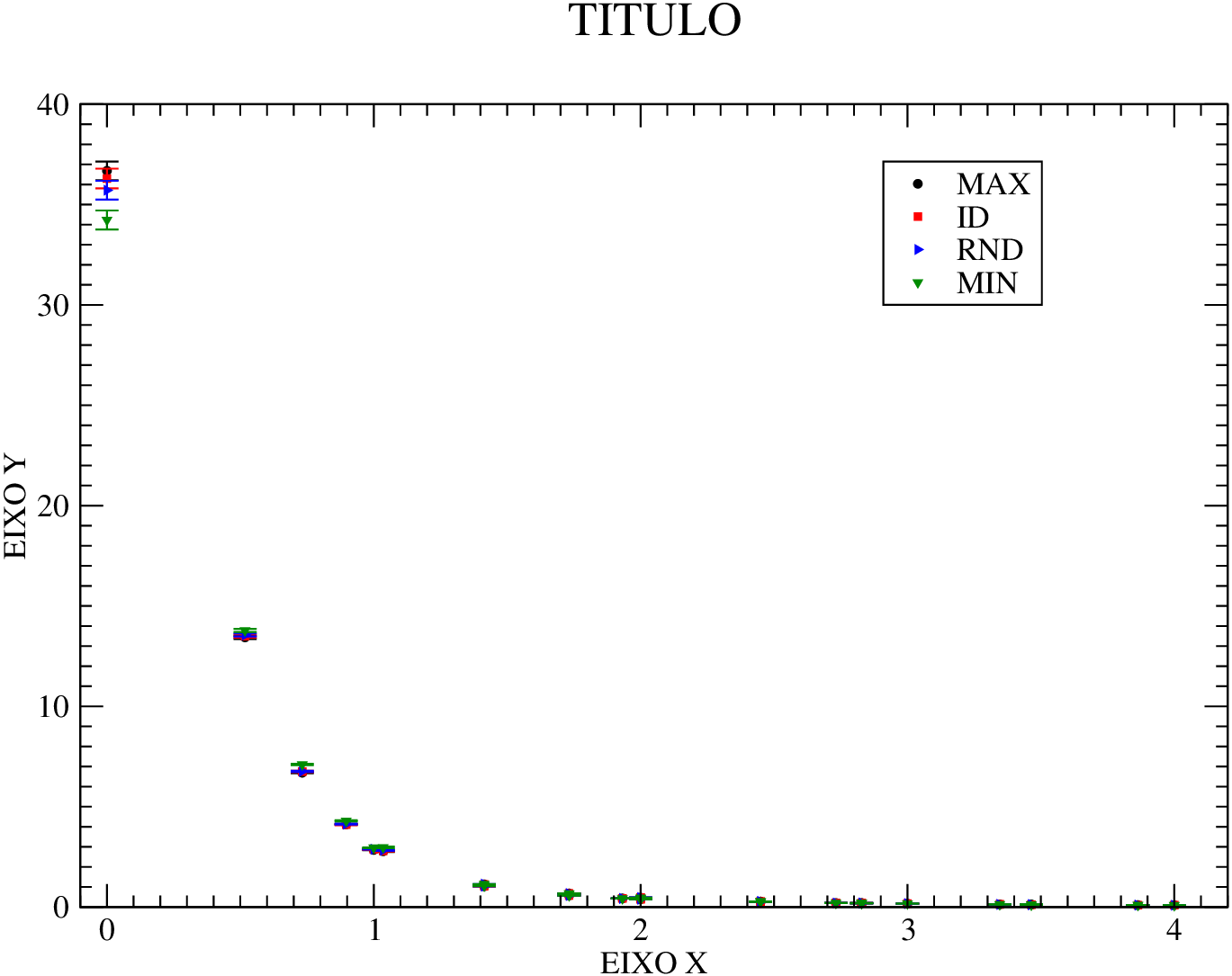}
  \end{minipage} }\\
\caption{Bare gluon propagator. Statistical errors were computed using the
jackknife procedure.}
\label{Fig_Dq2}
\end{figure}
\end{center}

\subsection{Bare Gluon Propagator}

\begin{table}[t]
\begin{center}
\begin{tabular}{||l@{\hspace{0.3cm}}l@{\hspace{0.3cm}}||rrrr||}
\hline\hline
 $n_\mu$  &  $|q|$  & \multicolumn{4}{|c||}{$D(q^2)$} \\
            &         & MAX & ID & RND & MIN \\
\hline
 $(0,0,0,0)$  & $0.0000$  &  $36.68(46)$   &   $36.30(49)$    &   
                             $35.71(47)$   &   $34.23(47)$   \\
 $(1,0,0,0)$  & $0.5176$  &  $13.436(79)$  &   $13.496(79)$   &   
                             $13.556(75)$  &   $13.780(77)$  \\
 $(2,0,0,0)$  & $1.0000$  &   $2.848(18)$  &    $2.873(17)$   &    
                              $2.881(19)$  &    $2.955(17)$  \\
 $(3,0,0,0)$  & $1.4142$  &   $1.0363(64)$ &    $1.0397(64)$  &    
                              $1.0415(63)$ &    $1.0566(61)$ \\
 $(4,0,0,0)$  & $1.7320$  &   $0.5769(34)$ &    $0.5772(30)$  &    
                              $0.5771(34)$ &    $0.5767(34)$ \\
 $(5,0,0,0)$  & $1.9319$  &   $0.4278(24)$ &    $0.4280(25)$  &    
                              $0.4293(24)$ &    $0.4316(25)$ \\
 $(6,0,0,0)$  & $2.0000$  &   $0.3892(32)$ &    $0.3868(32)$  &    
                              $0.3878(30)$ &    $0.3840(32)$ \\
\hline
 $(1,1,0,0)$  &  $0.7320$  &  $6.693(32)$   &  $6.752(35)$   &  
                              $6.760(35)$    &  $7.100(38)$    \\
 $(2,2,0,0)$  &  $1.4142$  &  $1.1303(61)$  &  $1.1266(61)$  &  
                              $1.1349(56)$   &  $1.1422(55)$   \\
 $(3,3,0,0)$  &  $2.0000$  &  $0.4377(20)$  &  $0.4390(20)$  &  
                              $0.4398(21)$   &  $0.4401(21)$   \\
 $(4,4,0,0)$  &  $2.4495$  &  $0.2635(13)$  &  $0.2636(12)$  &  
                              $0.2637(13)$   &  $0.2646(13)$   \\
 $(5,5,0,0)$  &  $2.7320$  &  $0.2026(10)$  &  $0.2020(10)$  &  
                              $0.2024(10)$   &  $0.2019(10)$   \\
 $(6,6,0,0)$  &  $2.8284$  &  $0.1866(13)$  &  $0.1863(12)$  &  
                              $0.1867(13)$   &  $0.1859(12)$   \\
\hline
 $(1,1,1,0)$  &  $0.8966$  &  $4.123(27)$   &  $4.109(26)$   &  
                              $4.131(27)$   &  $4.295(27)$    \\
 $(2,2,2,0)$  &  $1.7320$  &  $0.6725(41)$  &  $0.6737(44)$  &  
                              $0.6693(41)$  &  $0.6736(42)$   \\
 $(3,3,3,0)$  &  $2.4495$  &  $0.2734(16)$  &  $0.2747(16)$  &  
                              $0.2723(16)$  &  $0.2761(16)$   \\
 $(4,4,4,0)$  &  $3.0000$  &  $0.1681(10)$  &  $0.1688(10)$  &  
                              $0.1692(10)$  &  $0.1709(10)$   \\
 $(5,5,5,0)$  &  $3.3461$  &  $0.13105(75)$ &  $0.13264(75)$ &  
                              $0.13147(74)$ &  $0.13156(79)$   \\
 $(6,6,6,0)$  &  $3.4641$  &  $0.1216(10)$  &  $0.1222(10)$  &  
                              $0.1208(10)$  &  $0.1230(11)$   \\
\hline
 $(1,1,1,1)$  &  $1.0353$  &  $2.775(33)$   &  $2.795(33)$   &   
                              $2.831(34)$   &   $2.972(38)$   \\
 $(2,2,2,2)$  &  $2.0000$  &  $0.4674(53)$  &  $0.4664(56)$  &   
                              $0.4730(54)$  &   $0.4674(56)$   \\
 $(3,3,3,3)$  &  $2.8284$  &  $0.2018(25)$  &  $0.1993(25)$  &   
                              $0.1995(25)$  &   $0.1967(24)$   \\
 $(4,4,4,4)$  &  $3.4641$  &  $0.1238(16)$  &  $0.1233(15)$  &   
                              $0.1228(16)$  &   $0.1251(15)$   \\
 $(5,5,5,5)$  &  $3.8637$  &  $0.0982(12)$  &  $0.0972(12)$  &   
                              $0.0981(10)$  &   $0.0965(11)$   \\
 $(6,6,6,6)$  &  $4.0000$  &  $0.0894(14)$  &  $0.0904(15)$  &   
                              $0.0899(15)$  &   $0.0898(14)$   \\
\hline\hline
\end{tabular}
\caption{Bare gluon scalar function. The numbers in parentheses are 
the statistical errors, computed using the jackknife procedure, on the last
digits of $D(q^2)$.}
\label{Tab_D_q}
\end{center}
\end{table}

The scalar function $D(q^2)$, computed according to equations (\ref{Dq2}) and 
(\ref{Dq20}), after averaging over equivalent momenta\footnote{For example, 
for each gauge configuration the quoted value for momenta $(1,0,0,0)$ is the 
average over $(1,0,0,0)$, $(0,1,0,0)$, $(0,0,1,0)$ and $(0,0,0,1)$ values. 
Similarly, for $(1,1,0,0)$ a $Z_4$ average is performed, etc.}, 
is shown in figure
\ref{Fig_Dq2} as function of $\hat{q}$ and as function of $q$. The figures 
include $D(q^2)$ as function of momenta of type $(q,0,0,0)$, $(q,q,0,0)$, 
$(q,q,q,0)$ and $(q,q,q,q)$ for all available $q$ in our lattice. 
The figures for $D(q^2)$ are reported in table \ref{Tab_D_q}. 
From now on, unless stated clearly, 
we will consider only the data refering to $D(q^2)$ as function of $q$. 

Figure \ref{Fig_Dq2} and table  \ref{Tab_D_q} show that, for the gluon
propagator, the effect of Gribov copies is small and visible for 
the smallest momenta. Indeed, comparing the different gluon propagators to the
MAX propagator, it comes that, within one standard deviation, the ID 
propagator agrees with the $D(q^2)$ MAX for almost all the momenta considered.
The exception being $D(q^2)$ for the momenta associated to 
$n_\mu = (5,5,5,0)$, compatible with the MAX value only within two standard
deviations. Note that only the ID and MAX values agree for the infrared regime.
The RND propagator agrees, within one standard deviation, with the MAX 
propagator for all momenta but the zero momenta. 
The zero momenta RND propagator agrees 
with the MAX $D(0)$ only within two standard deviations. The strongest
deviation from the MAX propagator occurs when $D(q^2)$ is computed using
the smallest of the maximum of $F_U$. The MIN propagator agrees, within one
standard deviation, with MAX for momenta $|q| \ge 1.7320$ for momenta of type
$(q,0,0,0)$, $|q| \ge 2.000$ for $(q,q,0,0)$, $|q| \ge 3.3461$ for momenta 
$(q,q,q,0)$ and $|q| \ge 3.4641$ for $(q,q,q,q)$ momenta. 
For smaller momenta the differences between the
$D(q^2)$ values can achieve six standard deviations. Indeed, the agreement
between the MIN and MAX values quoted in the table are: six standard
deviations for $n_\mu = (1,1,0,0)$; four standard deviations for
$n_\mu = (1,1,1,0)$ and $(2,0,0,0)$; three standard deviations for 
$n_\mu = (0,0,0,0)$, $(1,0,0,0)$ and $(1,1,1,1)$; two standard deviations for 
$n_\mu = (2,2,0,0)$, $(3,0,0,0)$, $(3,3,3,3)$ and $(4,4,4,0)$. 
The lattice data shows clearly that Gribov
copies change the low momenta ($q < 1.7320$) components of the
gluon propagator.

\begin{center}
\psfrag{EIXO X}{$q$}
\psfrag{EIXO Y}{$D/D_{MAX}$}
\psfrag{TITULO}{}
\begin{figure}[t]
\includegraphics[origin=c,scale=0.45]{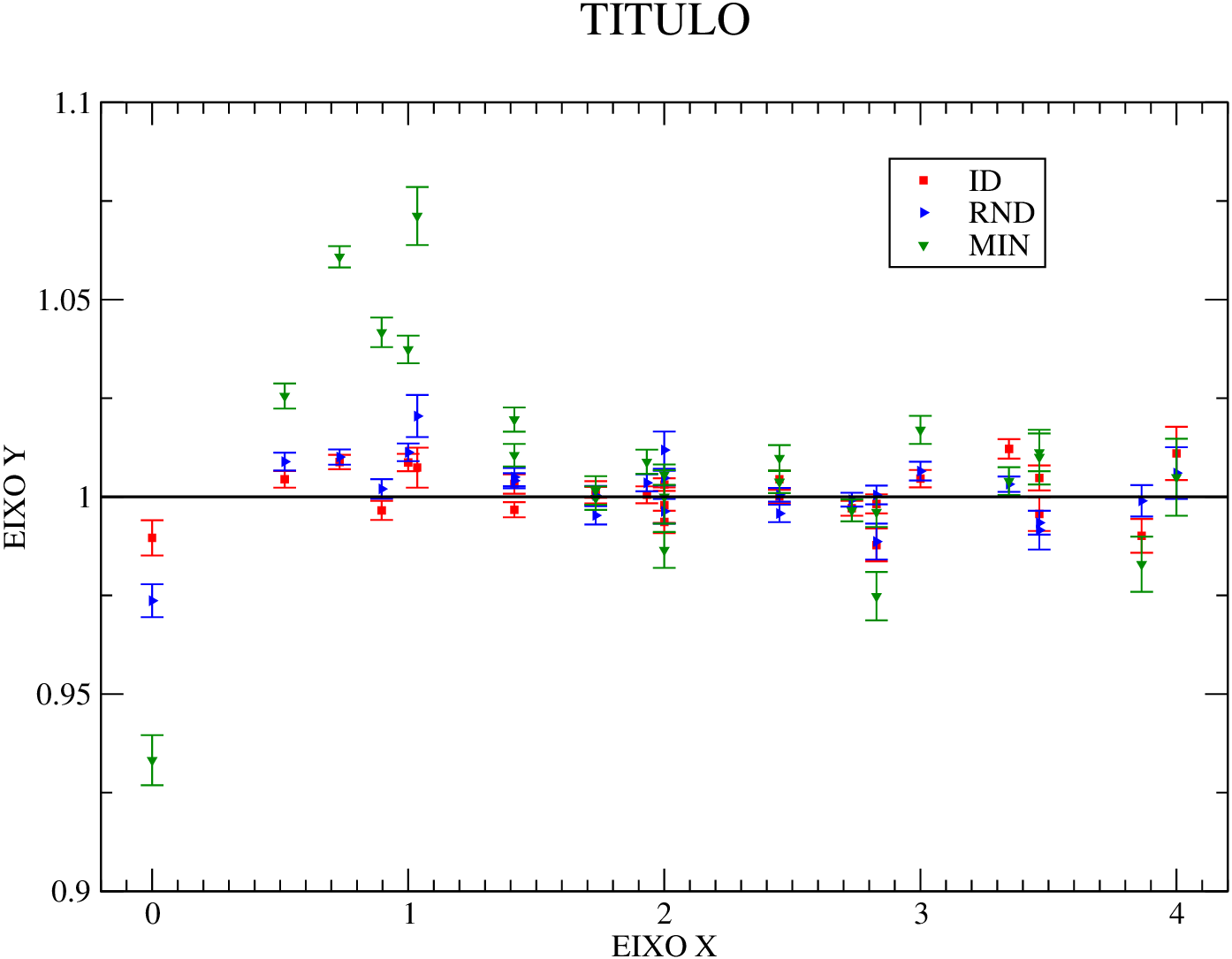}
\caption{$D(q^2) / D_{MAX} (q^2)$ as function of $q$ for ID, RND and MIN
propagators. Statistical errors were computed using the bootstrap method. The
quoted errors correspond to a 68\% confidence limit of the distributions 
obtained from 5000 bootstrap samples.}
\label{Fig_DDmax}
\end{figure}
\end{center}

For zero momentum, the largest propagator occurs when the configurations are
gauge fixed to the fundamental modular region. 
The absolute difference between the MIN, RND and ID to the MAX zero momenta 
propagator central values are 6.7\%, 2.6\% and 1\%, respectively. These 
numbers can be read as an order of magnitude of the maximal change on the 
gluon propagator due to Gribov copies.
For the other momenta, it is not always true that the largest value of 
$D(q^2)$ is associated to the MAX propagator. This can be seen in figure 
\ref{Fig_DDmax}. 

Figure \ref{Fig_DDmax} suggests that the ratio between 
the propagators to the MAX propagator is a function of $q$, that converges to 
one for the larger momenta. Moreover, the figure shows clearly that
the MIN propagator is different from the MAX propagator for momenta smaller 
than $q \sim 1.7$. From figure \ref{Fig_DDmax} one can quantify again
the change on the gluon propagator due to Gribov copies. 
For the MIN propagator, the effect of Gribov copies is, at most, a factor of
5-10\%. For the RND and ID propagators, the effect of Gribov copies is not so
dramatic (a factor smaller than 5\%).

\begin{center}
\begin{figure}[t]
  \subfigure[$D(q^2)  D_{\mbox{MAX}} (0) / D(0)$ as function of $q$ 
             for all momenta.]{\label{Fig_D_sca0}
  \begin{minipage}[b]{\textwidth}
    \psfrag{EIXO X}{$q$}
    \psfrag{EIXO Y}{$D(q^2) ~ D_{\mbox{MAX}} (0) / D(0)$}
    \psfrag{TITULO}{}
    \centering
    \includegraphics[origin=c,scale=0.43]{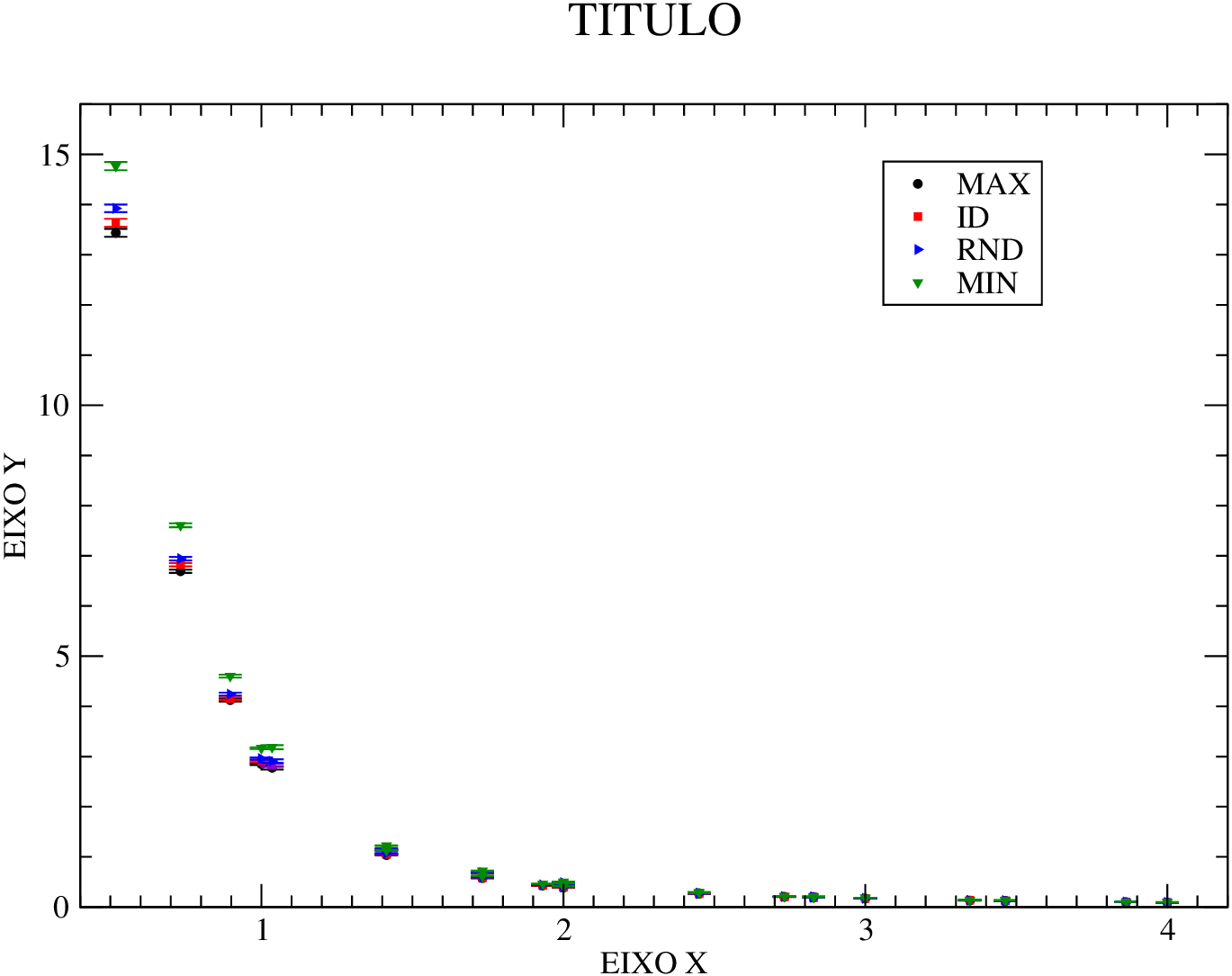}
  \end{minipage} } \\

  \subfigure[$D(q^2)  D_{\mbox{MAX}} (0) / D(0)$ as function of $q$ 
             for the larger momenta.]{ \label{Fig_D_sca0_zoom}
  \begin{minipage}[b]{\textwidth}
    \psfrag{EIXO X}{$q$}
    \psfrag{EIXO Y}{$D(q^2) ~ D_{\mbox{MAX}} (0) / D(0)$}
    \psfrag{TITULO}{}
    \centering
    \includegraphics[origin=c,scale=0.43]{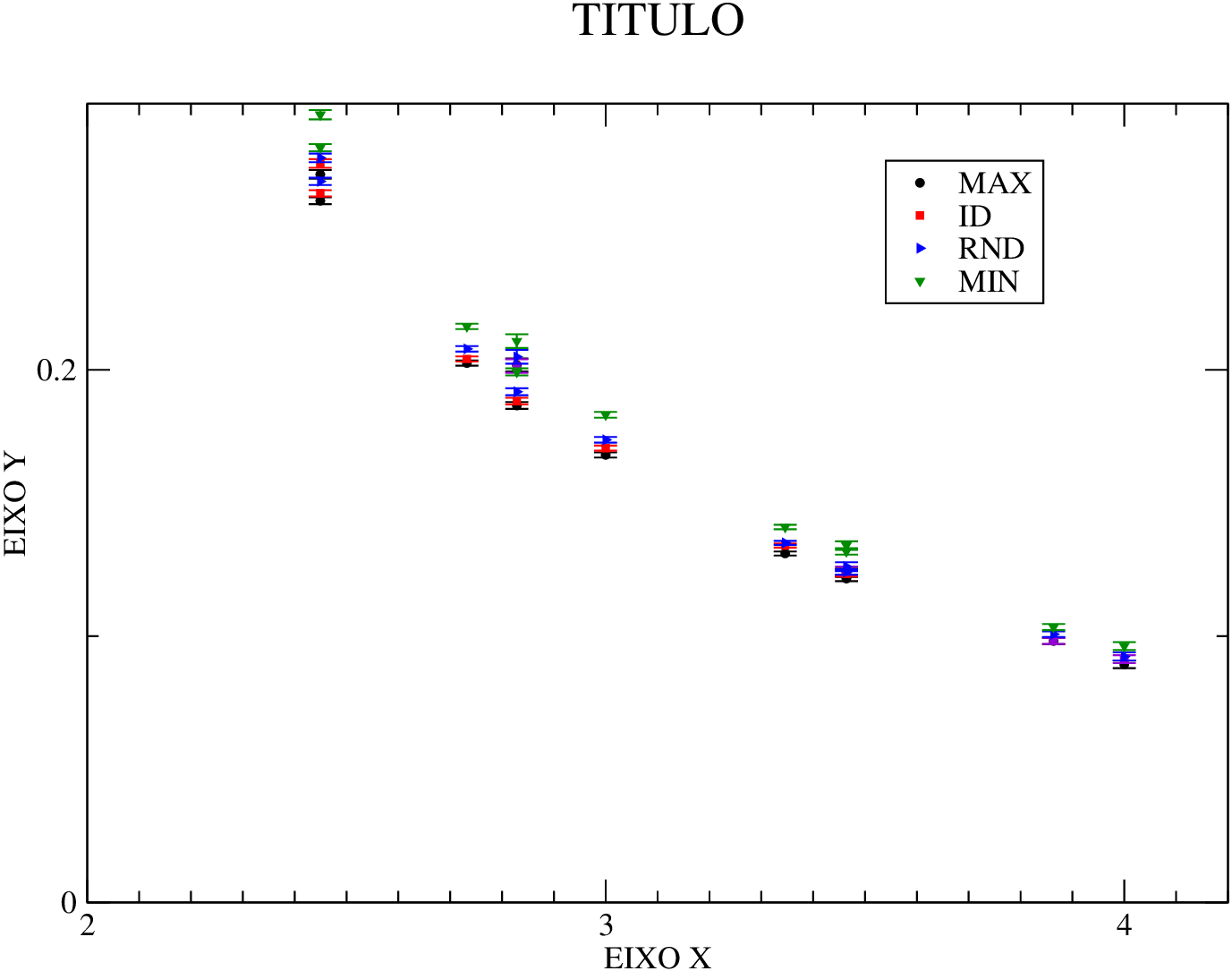}
  \end{minipage} }\\
\caption{Scaled gluon propagator. Statistical errors were computed using the 
bootstrap method. Statistical errors were computed using the jackknife 
procedure.}
\label{Fig_Dsca0}
\end{figure}
\end{center}

\begin{center}
\psfrag{EIXO X}{$i$}
\psfrag{EIXO Y}{$\chi^2 / d.o.f.$}
\psfrag{TITULO}{}
\begin{figure}[t]
\includegraphics[origin=c,scale=0.45]{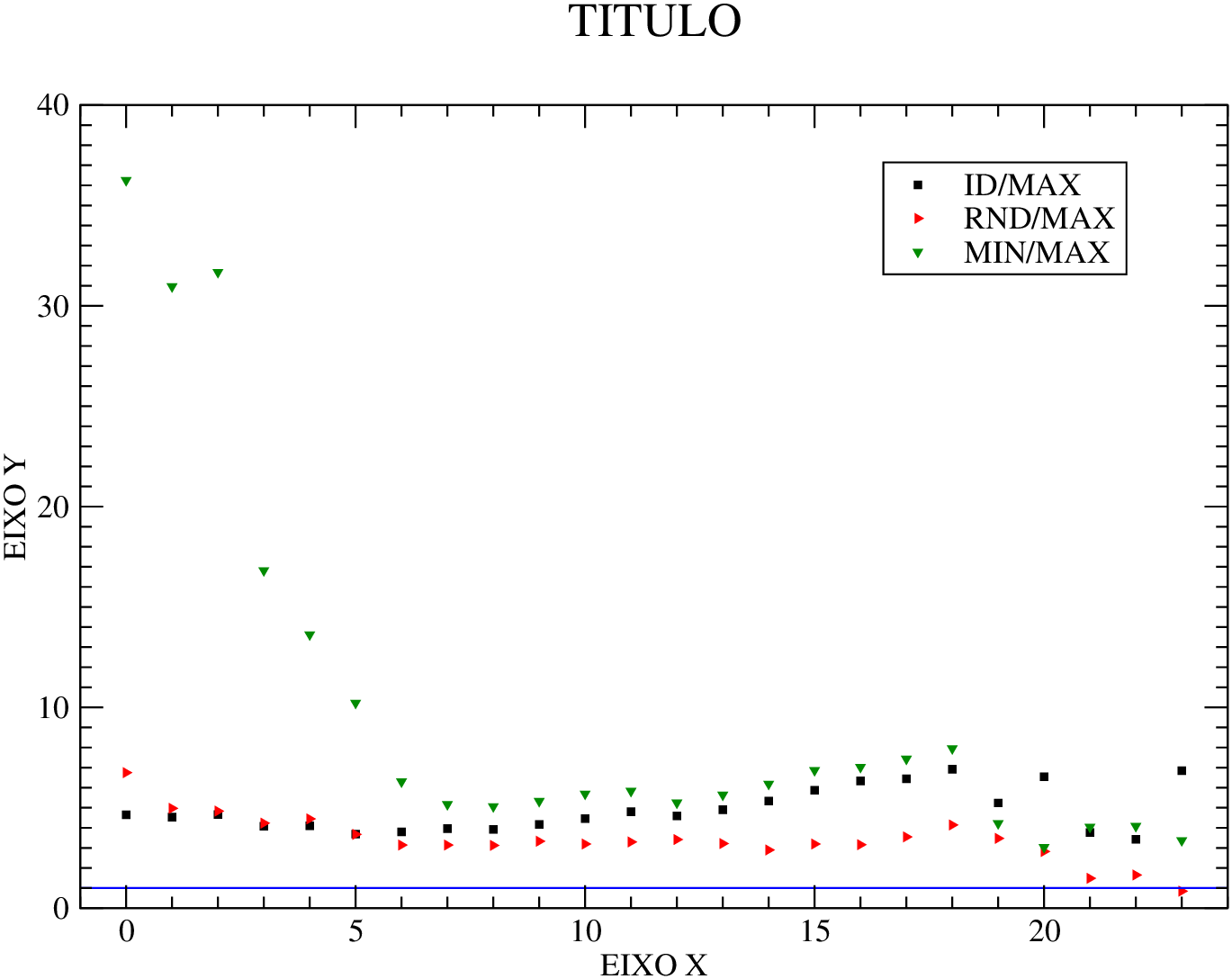}
\caption{$\chi^2/d.o.f.$ for the uncorrelated fits of $D(q^2) / D_{MAX} (q^2)$ 
to a constant.In the figure, $i$ is the number of lower momenta not considered
in the fit.}
\label{Fig_DDmax_fit}
\end{figure}
\end{center}

Figure \ref{Fig_DDmax} could suggest that the ratio between the
propagators to the MAX propagator would be a constant factor. 
To test this hypothesis,
in figure \ref{Fig_Dsca0} the propagators are ploted after rescaling the 
different gluon fields to reproduce the central value of the zero momenta
MAX scalar function. As seen in figure \ref{Fig_Dsca0}, the propagators differ
by more than one standard deviation for certain momenta. If, instead of 
rescaling the gluon field to reproduce the zero momenta MAX scalar function,
the matching is done, for example, for $n_\mu = (6,0,0,0)$, then
$D(0) = 36.68 \pm 0.46$, $36.53 \pm 0.50$, $35.84 \pm 0.47$ and
$34.70 \pm  0.48$ for the MAX, ID, RND and MIN propagators, respectively.
The MAX, ID and RND $D(0)$ are compatible within one standard deviation.
The
MIN $D(0)$ agrees with MAX value only within three standard deviations. In 
order to further test the hypothesis under discussion, $D / D_{MAX}$ 
was
fitted to a constant. 
No correlations were considered in the fits. The $\chi^2 /d.o.f.$ 
for these
fits are reported in figure \ref{Fig_DDmax_fit}. Although, in general,
the values of the $\chi^2 / d.o.f.$ decrease as one excludes more lower
momenta, they are always too high to conclude that the ID, RND and MIN
propagators differ, from the MAX propagator, by a multiplicative factor. In 
particular, the difference between MIN, ID and MAX propagators is clearly not 
a constant. The RND/MAX ratio is compatible with a constant for the largest 
momenta considered.

In conclusion, the analysis of the raw data for the bare gluon propagator
suggests that the effect of Gribov copies is small, but observable
(clearly, less than a 10\% factor) and is stronger for smaller momenta. 
Moreover, Gribov copies have almost no effect on the high momentum components
of the gluon propagator. The data reported in table \ref{Tab_D_q} shows 
that the effect of Gribov copies can be overcame if one multiplies the 
statistical errors by a factor of 2 to 3 for the smaller momenta 
($a q \le 1.73$). This doubling of the statistical error can be either, a
general property associated to the effect of Gribov copies, or a result due
to the limited statistics used here. Note that in the SU(2) study of
\cite{cucchieri97}, the number of configurations used for the larger lattices
($12^4$, $16^4$) and for the larger $\beta$ ($= 2.7$) was about half or less
than half of the configurations used in our simulation. The investigation of
$D(q^2) / D_{\mbox{MAX}} (q^2)$ shows that the propagators associated to the
Gribov copies named as ID, RND and MIN differ from the MAX propagator by more
than a constant factor.

\subsection{Gribov copies and gluon propagator models}

In the previous section, it was argued that the ID, RND and MIN propagators
do not differ from the MAX propagator by a constant factor. The question we
would like to investigate now being: is it possible to quantify the 
differences, due to Gribov copies, when modelling the gluon propagator? 
To try to answer this question, we will study the fit of $D(q^2)$ to a 
functional form. 

In \cite{Le99} a number of gluon propagator models were studied. 
Our simulation access a limited range of momenta and, certainly, finite space
and/or finite volume effects are no negligeable. Instead of 
performing a detailed study of several functional forms, we chose to 
investigate the model which, according to Leinweber \textit{et al} 
\cite{Le99}, describes better the lattice data. 

Let us assume that the scalar function is given by
\begin{equation}
 D(q^2) ~ = ~ Z ~ \Bigg[ ~ \frac{A \, M^{2 \alpha}}
                              {\left( q^2 \, + \, M^2 \right)^{1 + \alpha}}
                       ~ + ~
                       \frac{L(q^2,M^2)}
                            {q^2 \, + \, M^2 }   ~ \Bigg] ~ ,
\label{D_Lein}
\end{equation}
where
\begin{equation}
 L(q^2,M^2) ~ = ~ \Bigg[  ~ \frac{1}{2} ~ 
                            \ln\left[ (q^2 \, + \, M^2)
                                      (q^{-2} \, + \, M^{-2})  \right] 
                          ~ \Bigg]^{-d_D}
\end{equation}
is an infrared-regulated version of the one-loop logarithm correction to the
gluon propagator and, for pure gauge theories, $d_D = 13/22$. 

According to the results of the previous section, Gribov copies seem to change
the gluon propagator for the low energy momenta. Therefore, to measure 
such an effect we will consider three different types of uncorrelated fits. 
A fit to the highest momenta (UV-fit) using the following functional form
\begin{equation}
 D(q^2) ~ = ~ \frac{Z}{q^2} ~ \Bigg\{  ~ \frac{1}{2} ~ 
                                \ln\left( \frac{q^2}{\Lambda^2} \right) 
                          ~ \Bigg\}^{-d_D} ~ ,
\label{D_Lein_UV}
\end{equation}
a one-loop corrected perturbative gluon propagator. A fit to the lowest momenta
(IR-fit), assuming that
\begin{equation}
 D(q^2) ~ = ~ \frac{A M^{2 \alpha}}
                       {\left( q^2 \, + \, M^2 \right)^{1 + \alpha}}
\label{D_Lein_IR}
\end{equation}
and a fit of (\ref{D_Lein}) to all lattice data.

In order to compare our results with \cite{Le99}, we take their central values
for $a^{-1}$ at $\beta = 6.0$ and $\beta = 6.2$ and scale $a$ to $\beta = 5.8$
using the results of two-loop calculations. This procedure gives, 
respectively, $a^{-1} = 1.463$ GeV and $a^{-1} = 1.590$ GeV. The average of 
the two values being $a^{-1} = 1.53 \pm 0.06$ GeV ($a = 0.13$ fm).

\begin{center}
\psfrag{EIXO X}{$q$}
\psfrag{EIXO Y}{$q^2 D(q^2)$}
\psfrag{TITULO}{}
\begin{figure}[t]
\includegraphics[origin=c,scale=0.45]{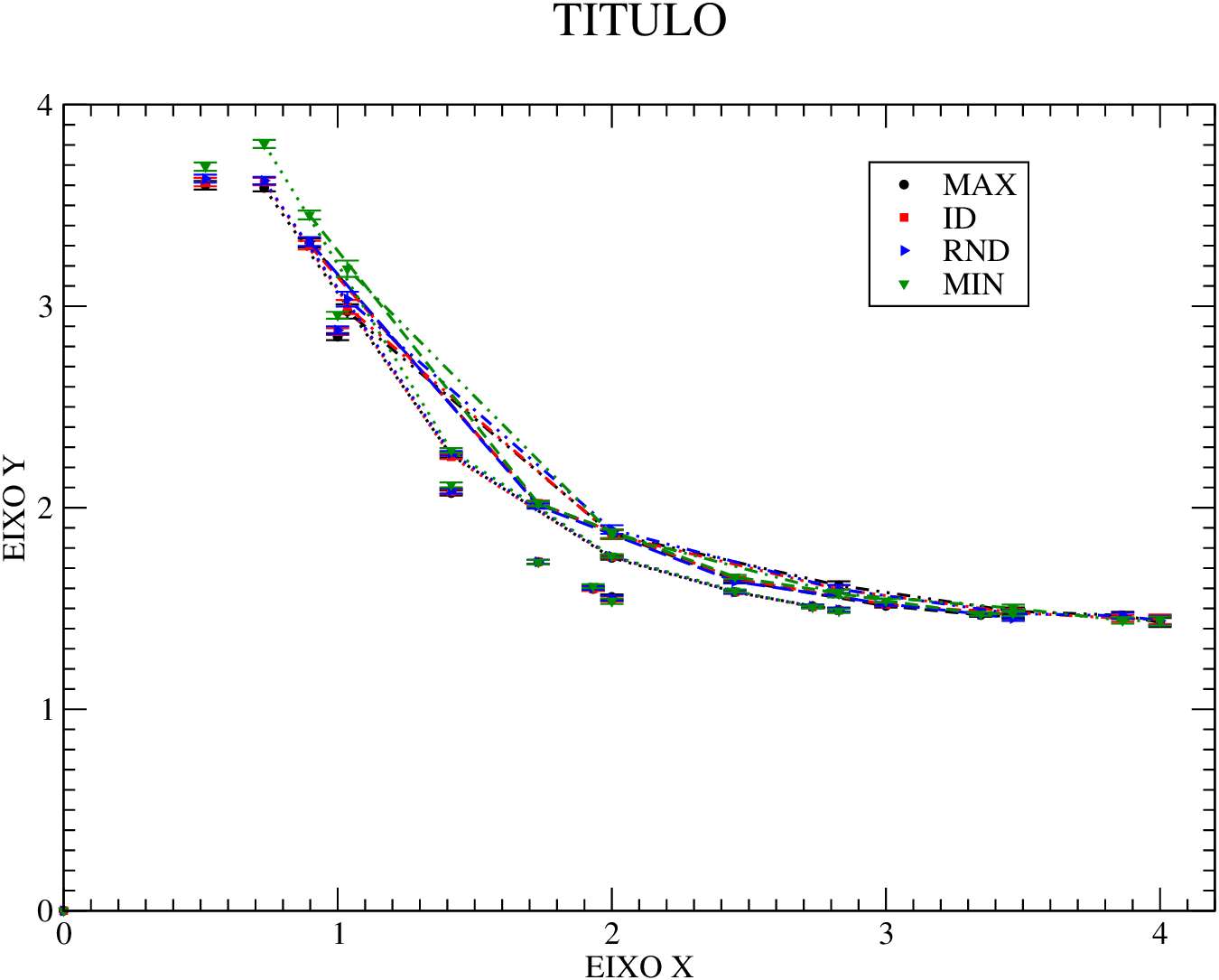}
\caption{$q^2 D(q^2)$ as function of $q$ for MAX, ID, RND and MIN
propagators. The points not connected by lines refer to $(q,0,0,0)$ momenta.
The points connect by dotted lines refer to $(q,q,0,0)$ momenta, the points
connected by dashed lines to $(q,q,q,0)$ momenta and the points connected by
dashed-dotted lines to $(q,q,q,q)$ momenta. Statistical errors were computed 
using the jackknife procedure.}
\label{Fig_q2D}
\end{figure}
\end{center}

Figure \ref{Fig_q2D} shows $q^2 D( q^2)$ as function of $q$ for all sets of 
gauge fixed configurations. The results for the different momenta shows that,
in our simulation, the finite space/volume effects are not
negligable - an effect of the order of 10\% from $(q,0,0,0)$ to
the other types of momenta. Since the different types of momenta have 
different finite space/volume effects, we will not include different types of
momenta in the fits. The exception being the IR fits.

\begin{table}[t]
\begin{center}
\begin{tabular}{||l@{\hspace{0.0cm}}|rrr@{\hspace{0.0cm}}||}
\hline\hline
              &   $Z$   &  $\Lambda$  & $\chi^2/d.o.f.$  \\
\hline\hline
MAX           &
   $1.473^{+11}_{-11}$  &  $0.8076^{+71}_{-69}$  &  $0.08$  \\
\hline
ID            &
   $1.4578^{+98}_{-98}$ &  $0.8181^{+65}_{-65}$  &  $0.48$   \\
\hline
RND           &
   $1.4620^{+97}_{-98}$ &  $0.8167^{+66}_{-67}$  &  $0.44$   \\
\hline
MIN           &
   $1.4243^{+98}_{-98}$ &  $0.8465^{+66}_{-62}$  &  $2.78$   \\
\hline\hline
\end{tabular}
\caption{Fits of the higher momenta of type $n_\mu = (n,0,0,0)$ to the one-loop
corrected perturbative gluon propagator (\ref{D_Lein_UV}). The fitting range
goes from $n = 3$ to $n = 6$. For larger fitting ranges, the $\chi^2/d.o.f.$ 
becomes too large ($> 18$). It is possible to fit the data using
a smaller fitting range ($n=4$ to $n=6$). However, we do not report the
figures because such a fit would have only one degree of freedom. 
Statistical errors 
were computed using the bootstrap method. The quoted errors correspond to a 
68\% confidence limit of the distributions obtained from 5000 bootstrap 
samples.}
\label{Tab_D_Lein_UV_i000}
\begin{tabular}{||l@{\hspace{0.0cm}}|rrr@{\hspace{0.0cm}}| 
                                     rrr@{\hspace{0.0cm}}||}
\hline\hline
              &  \multicolumn{3}{|c|}{$n_\mu = (2,2,0,0)$} 
              &  \multicolumn{3}{c||}{$n_\mu = (3,3,0,0)$} \\
              &   $Z$   &  $\Lambda$  & $\chi^2/d.o.f.$  
              &   $Z$   &  $\Lambda$  & $\chi^2/d.o.f.$  \\
\hline\hline
MAX           &
   $1.7846^{+61}_{-67}$  &  $0.7208^{+46}_{-41}$  &  $1.44$  &
   $1.861^{+19}_{-19}$   &  $0.659^{+15}_{-14}$  &  $0.14$  \\
ID            &
   $1.7867^{+51}_{-64}$  &  $0.7184^{+45}_{-35}$  &  $0.44$  &
   $1.823^{+17}_{-17}$   &  $0.688^{+13}_{-13}$  &  $0.15$  \\
RND           &
   $1.7776^{+57}_{-59}$  &  $0.7285^{+41}_{-38}$  &  $0.82$  &
   $1.829^{+18}_{-17}$   &  $0.686^{+13}_{-14}$  &  $0.27$  \\
MIN           &
   $1.7617^{+56}_{-55}$  &  $0.7413^{+39}_{-37}$  &  $0.52$  &
   $1.804^{+15}_{-17}$  &  $0.706^{+14}_{-12}$  &  $0.72$  \\
\hline\hline
\end{tabular}
\caption{Fits of the higher momenta of type $n_\mu = (n,n,0,0)$ to the one-loop
corrected perturbative gluon propagator (\ref{D_Lein_UV}). The fitting range
goes from $n=2$ or $3$ up to $n = 6$. Statistical errors were computed using
the bootstrap method. The quoted errors correspond to a 68\% confidence limit
of the distributions obtained from 5000 bootstrap samples.}
\label{Tab_D_Lein_UV_ii00}
\end{center}
\end{table}
\begin{table}[t]
\begin{center}
\begin{tabular}{||l@{\hspace{0.0cm}}|rrr@{\hspace{0.0cm}}||}
\hline\hline
              &   $Z$   &  $\Lambda$  & $\chi^2/d.o.f.$  \\
\hline\hline
MAX           &
   $2.100^{+37}_{-36}$   &  $0.534^{+27}_{-27}$  &  $0.78$  \\
ID            &
   $2.165^{+35}_{-40}$   &  $0.498^{+27}_{-24}$  &  $1.29$  \\
RND           &
   $2.150^{+40}_{-39}$   &  $0.499^{+27}_{-27}$  &  $0.09$  \\
MIN           &
   $2.092^{+36}_{-33}$  &  $0.555^{+25}_{-27}$  &  $0.80$  \\
\hline\hline
\end{tabular}
\caption{Fits of the higher momenta of type $n_\mu = (n,n,n,0)$ to the one-loop
corrected perturbative gluon propagator (\ref{D_Lein_UV}). The fitting range
goes from $3$ up to $n = 6$. For larger fitting ranges, the $\chi^2/d.o.f.$ 
becomes too large ($> 2$).  It is possible to fit the data using
a smaller fitting range ($n=4$ to $n=6$). However, we do not report the
figures because such a fit would have only one degree of freedom. 
Statistical errors were computed using
the bootstrap method. The quoted errors correspond to a 68\% confidence limit
of the distributions obtained from 5000 bootstrap samples.}
\label{Tab_D_Lein_UV_iii0}
\begin{tabular}{||l@{\hspace{0.0cm}}|rrr@{\hspace{0.0cm}}|
                                     rrr@{\hspace{0.0cm}}||}
\hline\hline
              &  \multicolumn{3}{|c|}{$n_\mu = (1,1,1,1)$}
              &  \multicolumn{3}{|c|}{$n_\mu = (2,2,2,2)$} \\
              &   $Z$   &  $\Lambda$  & $\frac{\chi^2}{d.o.f.}$
              &   $Z$   &  $\Lambda$  & $\frac{\chi^2}{d.o.f.}$  \\
\hline\hline
MAX           &
   $2.099^{+9}_{-11}$   &  $0.5944^{+47}_{-45}$ &  $0.39$  &
   $2.112^{+27}_{-31}$  &  $0.584^{+22}_{-19}$  &  $0.51$  \\
ID            &
   $2.076^{+8}_{-11}$   &  $0.6043^{+50}_{-41}$ &  $0.52$  &
   $2.102^{+27}_{-31}$  &  $0.584^{+24}_{-20}$  &  $0.64$  \\
RND           &
   $2.0756^{+79}_{-97}$ &  $0.6122^{+48}_{-42}$ &  $1.12$  &
   $2.064^{+26}_{-28}$  &  $0.621^{+22}_{-20}$  &  $1.48$  \\
MIN           &
   $2.0146^{+72}_{-93}$ &  $0.6516^{+46}_{-41}$ &  $1.20$  &
   $2.091^{+31}_{-30}$  &  $0.590^{+22}_{-23}$  &  $1.05$  \\
\hline\hline
\end{tabular}
\caption{Fits of the higher momenta of type $n_\mu = (n,n,n,n)$ to the one-loop
corrected perturbative gluon propagator (\ref{D_Lein_UV}). The fitting range
goes from $n=2$ or $3$ up to $n = 6$. Statistical errors were computed using
the bootstrap method. The quoted errors correspond to a 68\% confidence limit
of the distributions obtained from 5000 bootstrap samples.}
\label{Tab_D_Lein_UV_iiii}
\end{center}
\end{table}

The fits of the highest momenta to the asymptotic form (\ref{D_Lein_UV})
are reported in tables \ref{Tab_D_Lein_UV_i000} to \ref{Tab_D_Lein_UV_iiii}
for all types of momenta. The first point to remark is that
the gluon propagator scales perturbatively for $a q \ge \sqrt{2}$ for
momenta associated to $n_\mu = (n,0,0,0)$ and $n_\mu = (n,n,0,0)$, for
$a q \ge 2.450$ for $n_\mu = (n,n,n,0)$ momenta and for $a q \ge 1.035$ for 
$n_\mu = (n,n,n,n)$ momenta; i.e. the asymptotic form describes quite well
the lattice data for sufficiently large momenta. 
Perturbative scaling starts at momenta $q \sim 1.6 - 3.7$ GeV, a value 
compatible with the figure quoted in \cite{Le99}, 2.7 GeV. 

In what concerns the effect of Gribov copies at high momenta, the results
given in tables \ref{Tab_D_Lein_UV_i000} to \ref{Tab_D_Lein_UV_iii0} show
that, for the same data and fitting range, the MAX, ID and RND values are 
compatible within one standard deviation. For momenta associated to
$n_\mu = (n,n,n,n)$ and for the largest fitting 
range\footnote{In physical units, the fitting range includes momenta from 
1.6 GeV up to 6.1 GeV.}, the $Z$ and $\Lambda$ values are compatible within 
two standard deviations. On the other side, the MIN fitted parameters are
not compatible with the MAX figures; the exception being the fit to
$(q,q,q,0)$ momenta and the fit to the smallest fitting range reported in
table \ref{Tab_D_Lein_UV_iiii}.

In what concerns the stability of results, in general, the fitted parameters 
are not stable against a change in the fitting range. 
Probably, this is due to the limited number of different
momenta available in the
simulation. If one compares the results of the larger
fitting ranges where $Z$ and $\Lambda$ are compatible within one standard
deviation for the different types of momenta, it comes that $Z$ increases and
$\Lambda$ decreases as one goes from $n_\mu = (n,0,0,0)$ to 
$n_\mu = (n,n,n,n)$ by a factor of $\sim 1.4$. Such a large correction is an
indication of important finite space effects - remember that the lattice
spacing is $\sim 0.13$ fm. If one compares our values for $\Lambda$ with those 
reported in \cite{Le99}, the numbers given in tables \ref{Tab_D_Lein_UV_i000}
to \ref{Tab_D_Lein_UV_iiii} are, typically, larger than those reported by
Leinweber \textit{et al}. 

The discussion of the IR properties of the gluon propagator requires data for
small momenta. In our simulation one has only a limited access to the infrared
regime of QCD. This is a serious limitation to a proper investigation of the
low energy gluon propagator. Nevertheless, we have tried to find the 
combination of the smaller momenta which is well reproduced by
(\ref{D_Lein_IR}). Unfortunately, to achieve such a goal, we had to combine
different types of momenta. Below, we will show the results of such 
investigation. The reader should be aware of the physical meaning, or lack of
meaning, of the numbers reported here. We would like to remember that our main
goal is to see if there are differences, on the propagators, due to the choice
of Gribov copies.

The set of momenta associated\footnote{$q = 0$, $0.52$, $0.73$ and $1$ or, in 
physical units, $q = 0$, $0.80$, $1.12$ and $1.53$ GeV, respectively. Note
that the number of degrees of freedom for this fit is one.} to 
$n_\mu = (0,0,0,0)$, $(1,0,0,0)$, $(1,1,0,0)$ and $(2,0,0,0)$ 
is well described by the model function (\ref{D_Lein_IR}). The fitted 
parameters are reported in table \ref{Tab_D_Lein_IR} for the different
propagators. Although, the lattice data is well described by (\ref{D_Lein_IR}),
not all fitted parameters are compatible within one standard deviation. Indeed,
the MIN propagator values are not compatible with any of the other propagators.
The MAX and ID propagators all have the same $A$ parameter. The $A$ from the
RND fit is, within two standard deviations, compatible with the MAX figures.
In what concerns the gluon mass $M$, the MAX and ID values are compatible 
within one standard deviation but MAX and RND are compatible within three 
standard deviations. 
For the parameter $\alpha$, the MAX and ID values are compatible 
within one standard deviation but MAX and RND are compatible within two
standard deviations. Note that the gluon mass $M$ computed from the IR 
regime of QCD is not compatible, within one standard deviation, with the values
of $\Lambda$ from the UV regime - see tables \ref{Tab_D_Lein_UV_i000} to 
\ref{Tab_D_Lein_UV_iiii}. The values of $M$ and $\alpha$ for MAX are the
smallest figures in table \ref{Tab_D_Lein_IR}. From these fittings, one can 
quantify the effect due to Gribov copies as a two to three sigma effect on the
parameters.

\begin{table}[t]
\begin{center}
\begin{tabular}{||l@{\hspace{0.0cm}}|rrrr@{\hspace{0.0cm}}||}
\hline\hline
              &   $A$   &   $M$   &   $\alpha$   &   
                  $\chi^2/d.o.f.$  \\
\hline\hline
MAX           &
   $17.72^{+26}_{-24}$  &  $0.6947^{+62}_{-59}$  &
                           $1.278^{+23}_{-21}$   &
                           $0.048$                     \\
ID            &
   $18.19^{+29}_{-27}$  &  $0.7076^{+67}_{-68}$  &
                           $1.312^{+25}_{-24}$  &
                           $0.031$                     \\
RND           &
   $18.78^{+29}_{-27}$  &  $0.7237^{+68}_{-64}$  &
                           $1.363^{+26}_{-24}$   &
                           $1.032$                     \\
MIN           &
   $22.81^{+43}_{-44}$  &  $0.8189^{+94}_{-96}$  &
                           $1.675^{+37}_{-38}$  &
                           $1.561$                    \\
\hline\hline
\end{tabular}
\caption{The infrared propagator. Statistical errors were computed using the
bootstrap method. The quoted errors correspond to a 68\% confidence limit of 
the distributions obtained from 5000 bootstrap samples.}
\label{Tab_D_Lein_IR}
\end{center}
\end{table}

Finally, let us discuss the fittings of (\ref{D_Lein}) to all lattice data. 
The results of the fittings are reported in table \ref{Tab_D_Lein_i000} for
momenta $n_\mu = (n,0,0,0)$, in table \ref{Tab_D_Lein_ii00} for 
$n_\mu = (n,n,0,0)$, in table \ref{Tab_D_Lein_iii0} for $n_\mu = (n,n,n,0)$
and in table \ref{Tab_D_Lein_iiii} for  $n_\mu = (n,n,n,n)$ momenta. 
The $\chi^2/d.o.f.$ shows that, in general, the lattice data is well described 
by (\ref{D_Lein}). The exceptions are the fits to the MIN data, momenta
$n_\mu = (n,0,0,0)$, and ID propagator, momenta $n_\mu = (n,n,n,0)$.
For these two cases the $\chi^2/d.o.f.$ is quite large, meaning that
the lattice data is not described by (\ref{D_Lein}). 

To identify the effect of Gribov copies the different fittings are compared
for the same type of momenta. The data on tables \ref{Tab_D_Lein_i000} to 
\ref{Tab_D_Lein_iiii} shows that, for all types of momenta, the fitted
parameters for the MIN propagator are not compatible with the corresponding 
parameters for the MAX propagator. For momenta associated to
$n_\mu = (n,0,0,0)$, the ID and MAX propagators parameters are compatible
within one standard deviation. The RND and MAX $Z$ values are
compatible, within the same level of precision, the $\alpha$ and $M$ values 
are compatible within $2 \sigma$ and $A$ is compatible within three standard
deviations. For $n_\mu = (n,n,0,0)$ momenta,
ID and RND parameters are compatible with the MAX values only within two
standard deviations. The exception being the $\alpha$ from RND propagator,
which agrees with the MAX figures within $1 \sigma$.
For $n_\mu = (n,n,n,0)$, RND and MAX values are compatible
within two standard deviations. For the ID parameters, the $Z$ value is, 
within two standard deviation, compatible with the MAX value and all remaining
parameters are compatible within $1 \sigma$. For $n_\mu = (n,n,n,n)$, the
MAX, RND and ID fitted parameters are compatible within one standard
deviation; the $Z$ for the ID and MAX are compatible within $2 \sigma$.
Note that, in general, the MAX propagator has the larger $Z$ value and the 
smallest $A$, $M$ and $\alpha$. Again, like in the IR fits one can quantify
the effect due to Gribov copies as a two to three sigma effect. From the
fittings it is not possible to establish, clearly, which parameters are less
sensible to Gribov copies. Note that the fits to $n_\mu = (n,n,n,n)$ momenta,
although having large statistical errors and with the exception of the
MIN propagator, they do not distinguish the Gribov copies.

\begin{table}[t]
\begin{center}
\begin{tabular}{||l@{\hspace{0.0cm}}|rrrrr@{\hspace{0.0cm}}||}
\hline\hline
              &   $Z$   &   $A$   &   $M$   &   $\alpha$   &   
                            $\chi^2/d.o.f.$  \\
\hline\hline
MAX           &
   $1.581^{+11}_{-13}$  &  $12.63^{+29}_{-28}$   & 
                           $0.737^{+12}_{-12}$   &
                           $1.982^{+56}_{-56}$   &
                           $0.22$                     \\
ID            &
   $1.564^{+11}_{-11}$  &  $13.04^{+30}_{-27}$  &
                           $0.748^{+11}_{-11}$   &
                           $2.004^{+52}_{-52}$  &
                           $0.39$                     \\
RND           &
   $1.580^{+10}_{-10}$  &  $13.82^{+30}_{-27}$    &
                           $0.780^{+11}_{-11}$  &
                           $2.134^{+52}_{-48}$   &
                           $0.74$                     \\
MIN           &
   $1.559^{+09}_{-12}$   &  $15.69^{+32}_{-32}$    &
                           $0.841^{+12}_{-12}$  &
                           $2.320^{+54}_{-56}$  &
                           $3.45$                    \\
\hline\hline
\end{tabular}
\caption{Fits to all lattice data for momenta associated to 
$n_\mu = (n,0,0,0)$ to the functional form (\ref{D_Lein}). 
Statistical errors were computed using the bootstrap method. The
quoted errors correspond to a 68\% confidence limit of the distributions 
obtained from 5000 bootstrap samples.}
\label{Tab_D_Lein_i000}
\begin{tabular}{||l@{\hspace{0.0cm}}|rrrrr@{\hspace{0.0cm}}||}
\hline\hline
              &   $Z$   &   $A$   &   $M$   &   $\alpha$   &   
                            $\chi^2/d.o.f.$  \\
\hline\hline
MAX           &
   $1.8565^{+34}_{-41}$  &  $10.46^{+20}_{-18}$       &
                            $0.7283^{+76}_{-72}$  &
                            $1.990^{+32}_{-31}$     &
                            $1.15$                        \\
ID            &
   $1.8478^{+33}_{-39}$  &  $11.02^{+18}_{-18}$       &
                            $0.7493^{+72}_{-68}$   &
                            $2.061^{+28}_{-29}$     &
                            $0.12$                        \\
RND           &
   $1.8430^{+31}_{-38}$  &  $10.96^{+20}_{-18}$        &
                            $0.7524^{+77}_{-70}$  &
                            $2.046^{+30}_{-28}$     &
                            $0.45$                        \\
MIN           &
   $1.8055^{+32}_{-35}$  &  $13.94^{+20}_{-22}$       &
                            $0.8569^{+74}_{-78}$  &
                            $2.389^{+29}_{-30}$     &
                            $0.33$                        \\
\hline\hline
\end{tabular}
\caption{Fits to all lattice data for momenta associated to 
$n_\mu = (n,n,0,0)$, with $n$ from 0 to 6,
to the functional form (\ref{D_Lein}). Statistical errors were computed using
the bootstrap method. The
quoted errors correspond to a 68\% confidence limit of the distributions 
obtained from 5000 bootstrap samples.}
\label{Tab_D_Lein_ii00}
\end{center}
\end{table}
\begin{table}[t]
\begin{center}
\begin{tabular}{||l@{\hspace{0.0cm}}|rrrrr@{\hspace{0.0cm}}||}
\hline\hline
              &   $Z$   &   $A$   &   $M$   &   $\alpha$   &   
                            $\chi^2/d.o.f.$  \\
\hline\hline
MAX           &
   $1.9410^{+36}_{-48}$  &  $10.26^{+27}_{-27}$       &
                            $0.7371^{+94}_{-91}$  &
                            $2.018^{+39}_{-40}$     &
                            $1.71$                        \\
ID            &
   $1.9535^{+36}_{-47}$  &  $10.40^{+28}_{-25}$       &
                            $0.7484^{+97}_{-89}$   &
                            $2.071^{+41}_{-39}$     &
                            $2.46$                        \\
RND           &
   $1.9289^{+40}_{-45}$  &  $11.05^{+29}_{-29}$        &
                            $0.773^{+10}_{-10}$  &
                            $2.144^{+41}_{-42}$     &
                            $1.60$                        \\
MIN           &
   $1.9090^{+36}_{-42}$  &  $12.83^{+27}_{-26}$       &
                            $0.8460^{+95}_{-91}$  &
                            $2.355^{+37}_{-38}$     &
                            $0.68$                        \\
\hline\hline
\end{tabular}
\caption{Fits to all lattice data for momenta associated to 
$n_\mu = (n,n,n,0)$, with $n$ from 0 to 6,
to the functional form (\ref{D_Lein}). Statistical errors were computed using
the bootstrap method. The
quoted errors correspond to a 68\% confidence limit of the distributions 
obtained from 5000 bootstrap samples.}
\label{Tab_D_Lein_iii0}
\begin{tabular}{||l@{\hspace{0.0cm}}|rrrrr@{\hspace{0.0cm}}||}
\hline\hline
              &   $Z$   &   $A$   &   $M$   &   $\alpha$   &   
                            $\chi^2/d.o.f.$  \\
\hline\hline
MAX           &
   $2.018^{+12}_{-13}$  &  $10.38^{+73}_{-81}$       &
                           $0.756^{+24}_{-28}$  &
                           $2.15^{+10}_{-12}$     &
                           $0.48$                        \\
ID            &
   $1.993^{+11}_{-13}$  &  $11.07^{+75}_{-78}$       &
                           $0.780^{+24}_{-26}$   &
                           $2.22^{+10}_{-11}$     &
                           $0.44$                        \\
RND           &
   $2.001^{+10}_{-12}$  &  $10.42^{+72}_{-66}$        &
                           $0.764^{+24}_{-23}$  &
                           $2.117^{+98}_{-95}$     &
                           $1.12$                        \\
MIN           &
   $1.9239^{+70}_{-87}$  &  $14.30^{+74}_{-69}$       &
                            $0.896^{+21}_{-20}$  &
                            $2.535^{+89}_{-86}$     &
                            $0.57$                        \\
\hline\hline
\end{tabular}
\caption{Fits to all lattice data for momenta associated to 
$n_\mu = (n,n,n,n)$, with $n$ from 0 to 6,
to the functional form (\ref{D_Lein}). Statistical errors were computed using
the bootstrap method. The
quoted errors correspond to a 68\% confidence limit of the distributions 
obtained from 5000 bootstrap samples.}
\label{Tab_D_Lein_iiii}
\end{center}
\end{table}

From the analysis of tables \ref{Tab_D_Lein_i000} to \ref{Tab_D_Lein_iiii} 
one can check which parameters are robust against change of fitting momenta.
Indeed, looking only to the fundamental modular region propagators, it comes
that the overall normalization parameter $Z$ is a function of the type of
momenta considered. At $1 \sigma$, the different $Z$ values are not compatible
with each other. For the same level of precision, the $n_\mu = (n,n,0,0)$,
$n_\mu = (n,n,n,0)$ and $n_\mu = (n,n,n,n)$ fitted parameters which measures
the relative normalization of the infrared to ultraviolet propagator 
components, $A$, are compatible with each other. On the other hand,
$M$ and $\alpha$ parameters are robust against change of momenta. All four
values reported in the tables are compatible within one standard deviation.
Our results for $\alpha$ and $M$ are
\begin{center}
\begin{tabular}{l@{\hspace{0.2cm}}|r@{\hspace{0.4cm}}r@{\hspace{0.4cm}}
                                     r@{\hspace{0.4cm}}r}
\hline\hline
              &   $(q,0,0,0)$  &   $(q,q,0,0)$   &   
                  $(q,q,q,0)$  &   $(q,q,q,q)$ \\
\hline\hline
$\alpha$      &  $1.982^{+56}_{-56}$ &
                 $1.990^{+32}_{-31}$ &
                 $2.018^{+39}_{-40}$ &
                 $2.15^{+10}_{-12}$ \\
$M$ (MeV)     &  $1128^{+18}_{-18} \pm 44$ &
                 $1114^{+12}_{-11} \pm 44$ &
                 $1128^{+14}_{-14} \pm 44$ &
                 $1157^{+37}_{-43} \pm 45$ \\
\hline\hline
\end{tabular}
\end{center}
where the second error in $M$ represents the error in the lattice spacing. 
Curiously, these values are compatible, within one standard deviation with the
values quoted in \cite{Le99}, namely $\alpha = 2.2^{+0.1+0.2}_{-0.2-0.3}$ and
$M = (1020 \pm 100 \pm 25)$ MeV. Note that the values for $M$ and $\alpha$ 
quoted for the fittings to all lattice data are not, in general, compatible 
with the same parameters computed from the IR and UV fits. Probably, this is
due to using a relatively small lattice that does not enable a clear separation
between the low energy and high energy regimes of QCD.

\section{Discussion and Conclusions}

In this paper the problem of the Gribov copies in lattice QCD is addressed. 
To try to understand the role of Gribov copies in lattice QCD, the gluon
propagator was computed with 302 configurations for a $12^4$ lattice and for
$\beta = 5.8$ using the overrelaxed quasi-heat bath. 

\begin{center}
\begin{figure}[t]
  \subfigure[Zero momentum scalar function.]{\label{Fig_D0_evol}
  \begin{minipage}[b]{0.45\textwidth}
    \psfrag{EIXO X}{\begin{tiny} $\langle F_U \rangle$ \end{tiny}}
    \psfrag{EIXO Y}{$D(q^2)$}
    \psfrag{TITULO}{}
    \centering
    \includegraphics[origin=c,scale=0.25]{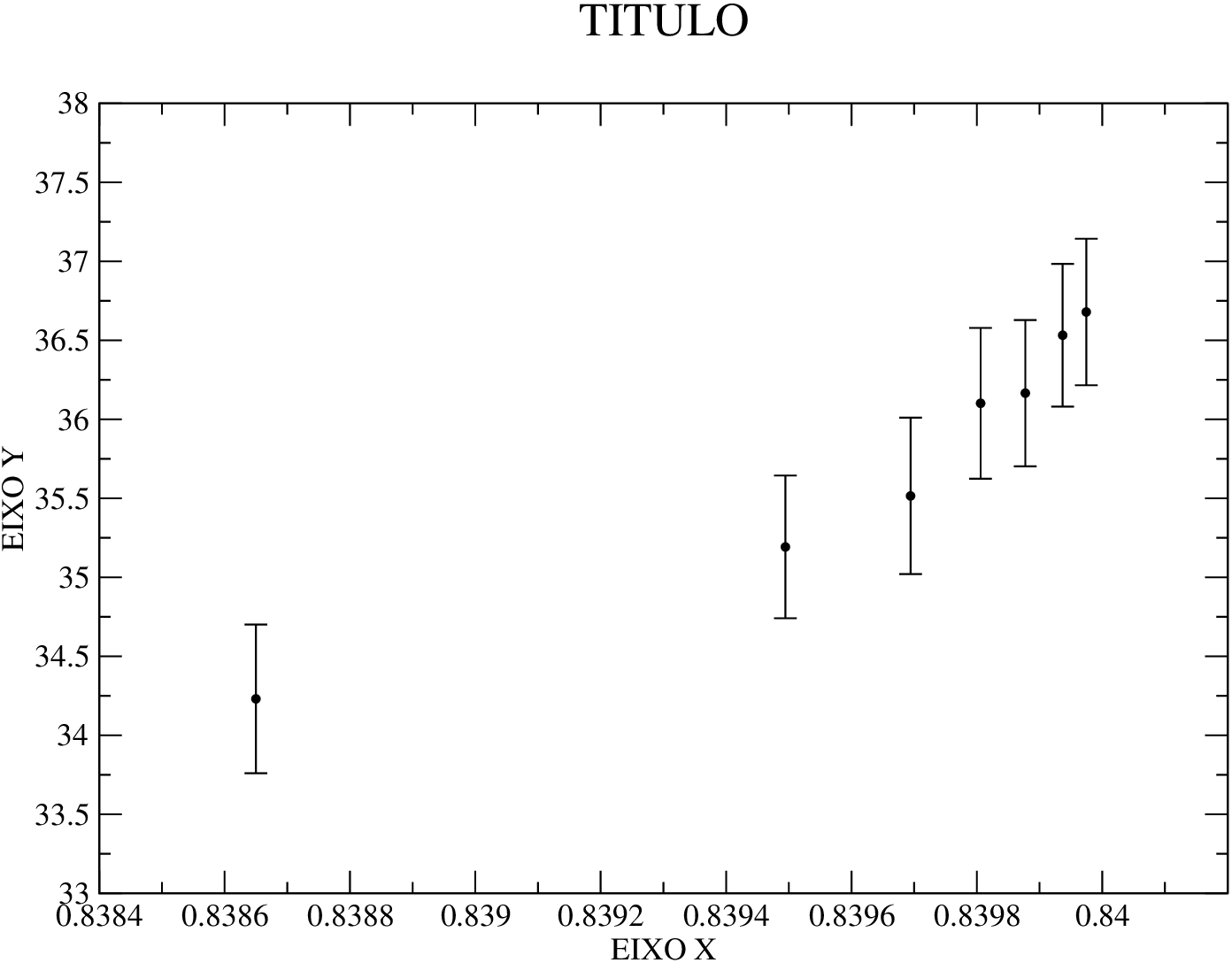}
  \end{minipage} } \hfill
  \subfigure[Scalar function associated to $n_\mu = (1,0,0,0)$.]{ \label{Fig_D1_evol}
  \begin{minipage}[b]{0.45\textwidth}
    \psfrag{EIXO X}{\begin{tiny} $\langle F_U \rangle$ \end{tiny}}
    \psfrag{EIXO Y}{$D(q^2)$}
    \psfrag{TITULO}{}
    \centering
    \includegraphics[origin=c,scale=0.25]{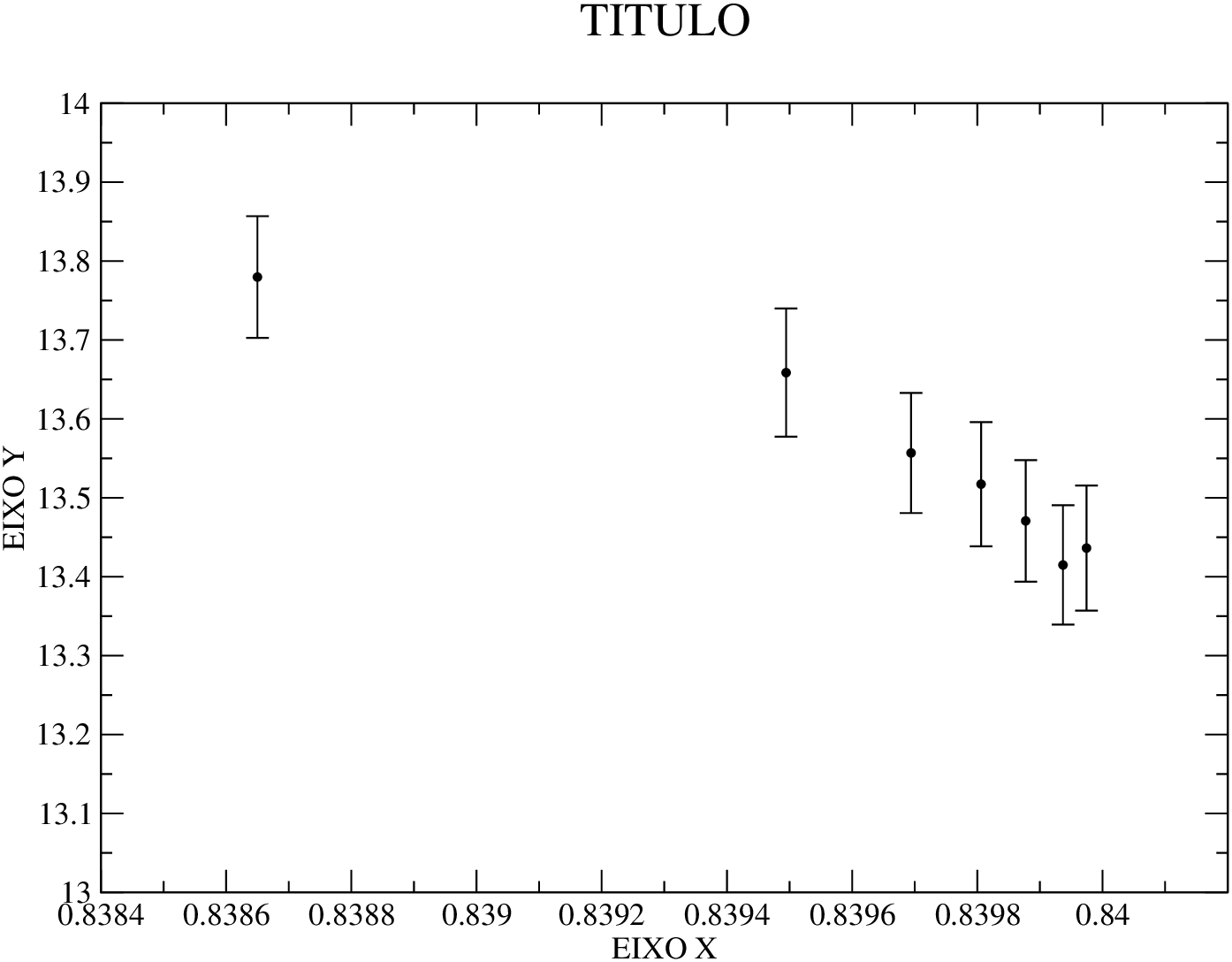}
  \end{minipage} }\\

  \subfigure[Scalar function associated to $n_\mu = (1,1,0,0)$.]{\label{Fig_D2_evol}
  \begin{minipage}[b]{0.45\textwidth}
    \psfrag{EIXO X}{\begin{tiny} $\langle F_U \rangle$ \end{tiny}}
    \psfrag{EIXO Y}{$D(q^2)$}
    \psfrag{TITULO}{}
    \centering
    \includegraphics[origin=c,scale=0.25]{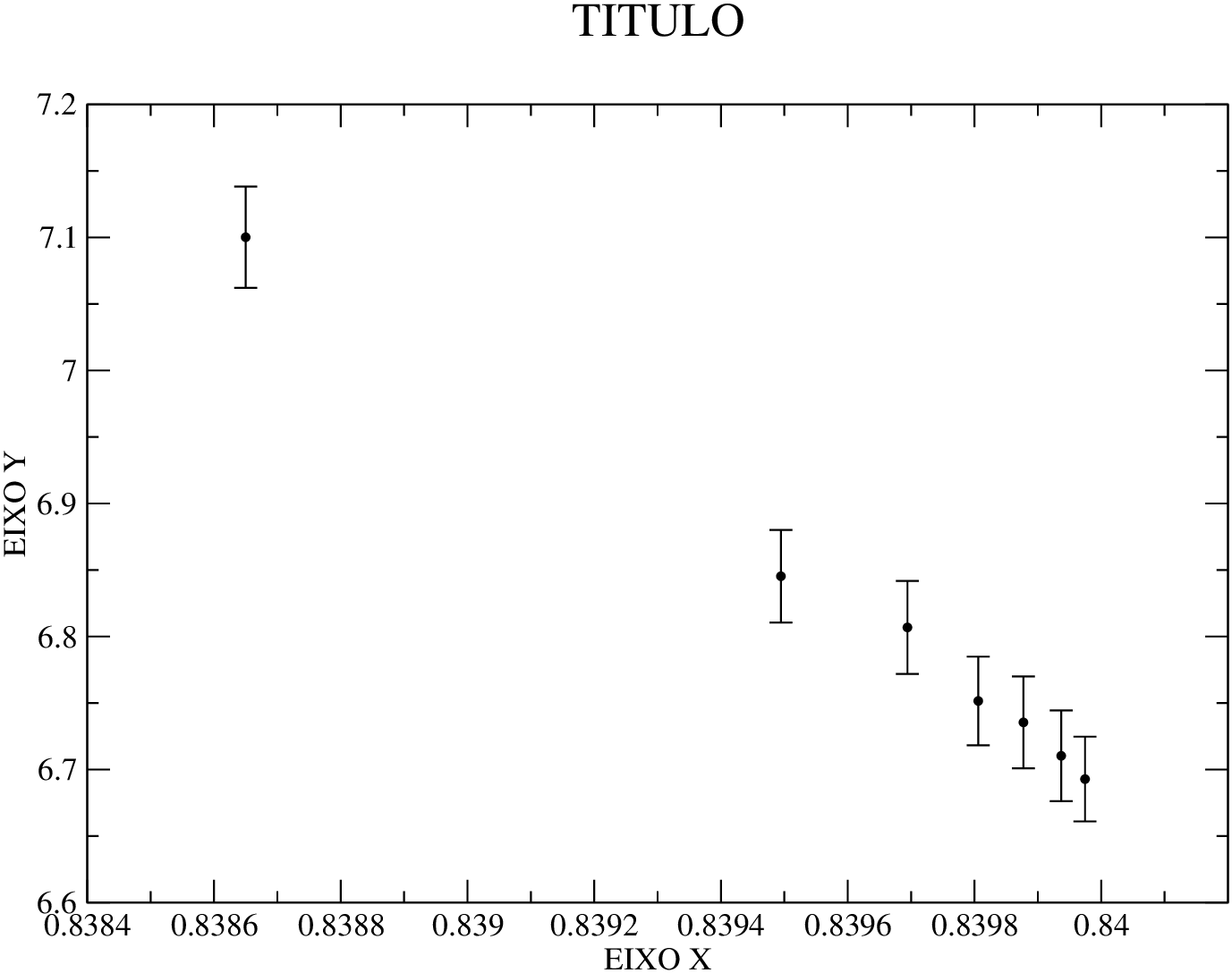}
  \end{minipage} } \hfill
  \subfigure[Scalar function associated to $n_\mu = (1,1,1,0)$.]{ \label{Fig_D3_evol}
  \begin{minipage}[b]{0.45\textwidth}
    \psfrag{EIXO X}{\begin{tiny} $\langle F_U \rangle$ \end{tiny}}
    \psfrag{EIXO Y}{$D(q^2)$}
    \psfrag{TITULO}{}
    \centering
    \includegraphics[origin=c,scale=0.25]{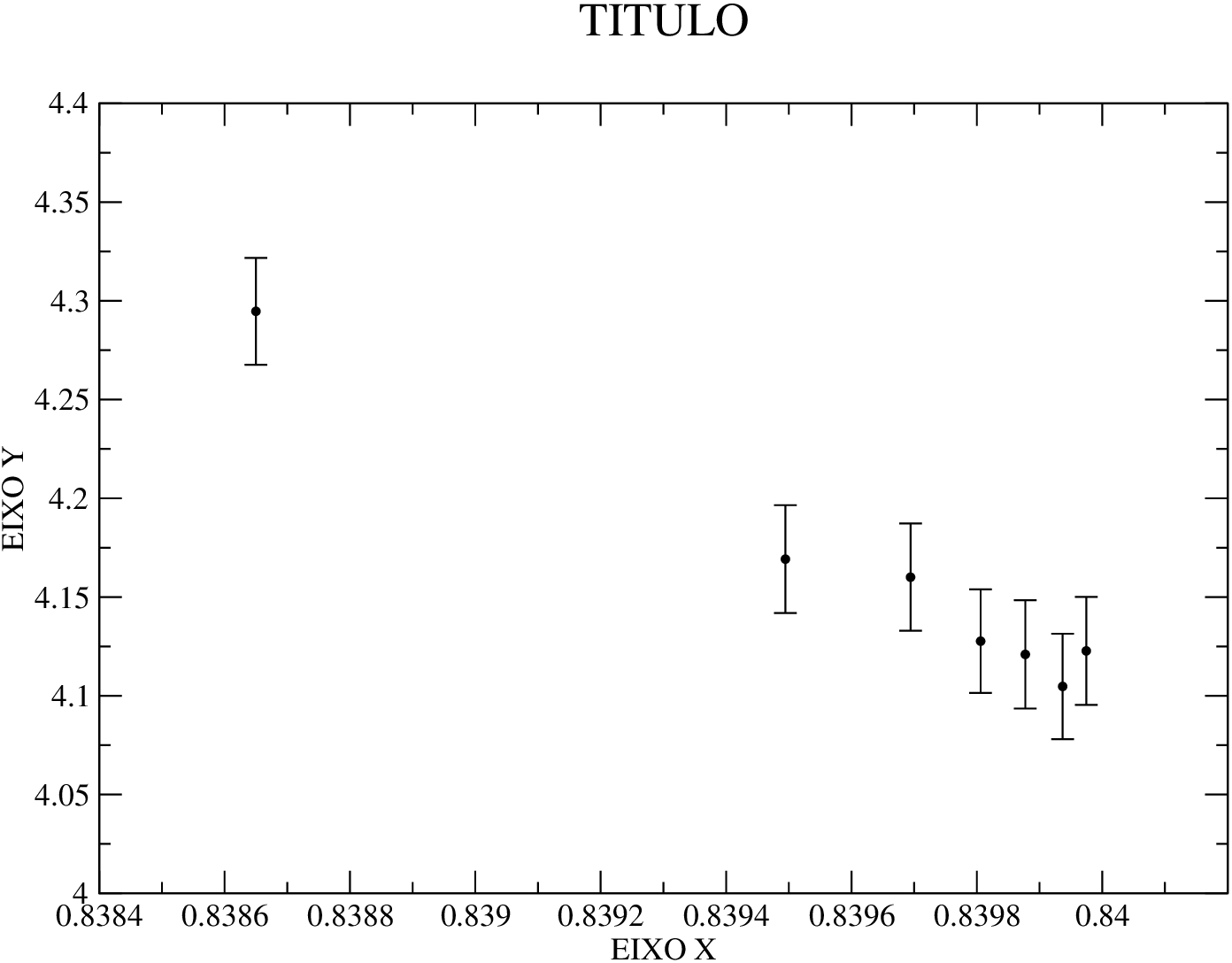}
  \end{minipage} }\\
\caption{Evolution of $D(q^2)$ with F. After ordering the gauge fixed
configurations associated with the sets MAX, ID, RND, MIN, RND1, 
RND2 and RND3 according to the $F_U$ value, the gluon scalar function is
computed picking always configurations within the same class of
values of $F_U$. Statistical errors were computed using the jackknife 
procedure.}
\label{Fig_D_evol}
\end{figure}
\end{center}

The analysis of the raw data shows that Gribov copies change only the low
momenta components of the gluon propagator. In our simulation, only for
momenta $aq < 1.7320$ ($q < 2.6$ GeV) there are significant 
differences between the MIN
and MAX propagators. The RND data is not compatible, within one standard
deviation, with the MAX propagator only for zero momentum. The study performed
here shows that, typically, the choice of different Gribov copies changes the
propagator in such a way that the figures become compatible within 
two-to-three standard deviations. Note, however, that for the patological case
of the MIN propagator the deviation relative to the MAX propagator, can be as
large as six standard deviations effects. This result seems to suggest that in
the study of the IR limit of the gluon propagator, the statistical errors 
should be multiplied by a factor of two or three in order to take into account 
possible deviations due to Gribov copies. If this is true for the statistical
accuracy of our study, this may not hold when larger statistics,
bigger lattices are considered.
That the Gribov copies change essentially the IR limit of the propagator can
be seen in figure \ref{Fig_D_evol}, where $D(q^2)$ is plotted against 
$\langle F_U \rangle$.

The properties observed for the raw scalar data are observed
when we model the lattice data. A difference, due to the choice of Gribov
copies, of up to two-to-three standard deviations is seen on the fitted
parameters. Note that this is observed even if the lowest energy momenta have
the largest absolute errors, i.e. there contribution to the $\chi^2$ is not
so relevant. To our mind, a deviation of this order of magnitude is, probably,
a good measure of the influence of the Gribov copies on the gluon propagator.

In what concerns the gluon propagator itself, the results of our simulation
for $M$ and $\alpha$ support the results quoted in a previous investigation
using much larger lattices \cite{Le99}. Remember that, for the MAX propagators,
these parameters are robust against a change on the type of fitted momenta.
It is interesting, that the lattice data supports quite well a gluon propagator
which, for large momenta, behaves like a 
massive vector with a mass of the order of the hadronic scale. Remember
that a massive gluon propagator, with a gluonic mass of the order of $1$ GeV,
has some phenomenological support \cite{Fi02}.

\vspace{0.5cm}

P. J. S. acknowledges financial support from the portuguese FCT. This
work was supported by grant SFRH/BD/10740/2002.

\end{document}